\documentclass[%
aip,
jcp,%
amsmath,amssymb,
reprint%
,longbibliography,
]{revtex4-1}

\usepackage{amsmath}%author-numerical,%
\usepackage{amsfonts}
\usepackage{amssymb,bm}
\usepackage{graphicx}
\usepackage{physics}
\usepackage{upgreek}
\usepackage[outercaption]{sidecap}   
\usepackage{xcolor}
\usepackage{hyperref}
\hypersetup{colorlinks=true,
	linkcolor=magenta,
	allcolors=magenta}
\usepackage{newtxtext}
\usepackage{newtxmath}
\usepackage[version=4]{mhchem}
\usepackage{blkarray}

\DeclareMathAlphabet{\pazocal}{OMS}{zplm}{m}{n}
\newcommand{\pP}{\ensuremath{\pazocal{P}}}

\newcommand{\pL}{\ensuremath{\pazocal{L}}}
\newcommand{\pK}{\ensuremath{\pazocal{K}}}
\newcommand{\pR}{\ensuremath{\pazocal{R}}}
\newcommand{\pG}{\ensuremath{\pazocal{G}}}

\newcommand{\sys}{\ensuremath{\mathrm{s}}}

\newcommand{\bath}{\ensuremath{\mathrm{b}}}

\renewcommand{\op}[1]{\ensuremath{\hat{#1}}}

\newcommand{\dla}{\langle\!\langle}
\newcommand{\dra}{\rangle\!\rangle}

\newcommand{\LE}{\mathrm{LE}}
\newcommand{\CT}{\mathrm{CT}}
\newcommand{\GS}{\mathrm{GS}}
\newcommand{\EM}{\mathrm{EM}}
\newcommand{\etal}{\textit{et al.}}
\newcommand{\Chla}{Chl\textit{a}}
\newcommand{\Chlb}{Chl\textit{b}}
\newcommand{\chla}[1]{\textit{a}#1}

\begin{document}

%\title{A hybrid hierarchical equations of motion method for simulating coupled electron transfer and exciton dynamics in light harvesting complexes}
\title{Coupled charge and energy transfer dynamics in light harvesting complexes from a hybrid hierarchical equations of motion approach}
%\title{Excitation quenching in light harvesting complexes from a hybrid hierarchical equations of motion approach}
\author{Thomas P. Fay\looseness=-1}
\email{tom.patrick.fay@gmail.com}
\affiliation{Department of Chemistry, University of California, Berkeley, CA 94720, USA\looseness=-1}
\author{David T. Limmer\looseness=-1}
\email{dlimmer@berkeley.edu}
\affiliation{Department of Chemistry, University of California, Berkeley, CA 94720, USA\looseness=-1}
\affiliation{Kavli Energy Nanoscience Institute at Berkeley, Berkeley, CA 94720, USA\looseness=-1}
\affiliation{Chemical Sciences Division, Lawrence Berkeley National Laboratory, Berkeley, CA 94720, USA\looseness=-1}
\affiliation{Materials Science Division, Lawrence Berkeley National Laboratory, Berkeley, CA 94720, USA\looseness=-1}
%\keywords{Chirality, electron transfer, spin, spintronics, biological electron transport}

%\begin{tocentry}
%\centering
%\includegraphics{toc-graphic-3.pdf}
%\end{tocentry}

\begin{abstract}
%Electron transfer forms an essential part of photosynthetic systems, such as in energy harvesting in reaction centers and in protective non-photochemical quenching in antenna complexes. 
We describe a method for simulating exciton dynamics in protein-pigment complexes, including effects from charge transfer as well as fluorescence. The method combines the hierarchical equations of motion, which are used to describe quantum dynamics of excitons, and the Nakajima-Zwanzig quantum master equation, which is used to describe slower charge transfer  processes. We study the charge transfer quenching in light harvesting complex II, a protein postulated to control non-photochemcial quenching in many plant species. Using our hybrid approach, we find good agreement between our calculation and experimental measurements of the excitation lifetime. Furthermore our calculations reveal that the exciton energy funnel plays an important role in determining quenching efficiency, a conclusion we expect to extend to other proteins that perform protective excitation quenching. This also highlights the need for simulation methods that properly account for the interplay of exciton dynamics and charge transfer processes.
\end{abstract}

	\maketitle
\section{Introduction}

Photosynthetic systems rely on both electronic excitation energy transfer and charge transfer processes to perform the reactions that sustain life on Earth.\cite{Mirkovic2017,Vinyard2013,Gorka2021,Blankenship2002,Croce2011} For example, excitation energy transfer (EET) and charge transfer (CT) play fundamental roles in reaction centers,\cite{Blankenship2002} where light energy from the Sun is harvested to drive chemical reactions. Charge transfer is also likely have an important photo-protective function in photosynthetic organisms,\cite{Goss2015,Park2019,Ruban2022,Pinnola2016,Ostroumov2020} by quenching excess excitation energy and preventing damage to photosynthetic systems. The importance of coupled charge and excitation energy transfer dynamics necessitates the development of theoretical methods to accurately and efficiently simulate them together. Here we develop a theory to study both processes using a hybrid approach that combines the hierarchical equation equations (HEOM) with quantum master equations (QME) to afford a computationally efficient method that is also accurate. 

Rapid excitation energy transfer has been studied extensively using a variety of methods, with HEOM emerging as a flexible and highly accurate approach for a large class of systems.\cite{Tanimura1989,Ishizaki2009a,Tanimura2020} The HEOM method has enabled the simulation of EET in photosynthetic complexes without invoking perturbation theory, enabling a balanced description of both incoherent F\"orster EET, and coherent excitonic EET, as well as transport dynamics intermediate between these two regimes.\cite{Ishizaki2009} %Additionally the HEOM method can capture superexchange\cite{Tanaka2009,Tanaka2010} and vibrational resonance effects\cite{Kreisbeck2014} in EET, all of which play a part in efficient energy transfer in light-harvesting complexes. 
%. This means that HEOM can capture both incoherent F\"orster EET, and coherent excitonic EET, as well as transport dynamics intermediate between these two regimes.\cite{Ishizaki2009} 
%
Although the HEOM method has been used extensively to study EET,\cite{Ishizaki2009a,Sarovar2010,Strumpfer2009,Ishizaki2010,Chen2011,Kreisbeck2011,Dijkstra2012,Kreisbeck2014,Schroter2015,Chan2018,Jankovic2020,Yan2021a} it has been used less in the study of combined EET and CT.\cite{Novoderezhkin2016} This is largely because charge transfer states typically couple much more strongly to the environment than local electronic excitations. Typical reorganization energies for CT processes are often in excess of $20 k_\mathrm{B} T$ at room temperature in polar environments, due to the large changes in charge density distributions on molecules involved in CT,\cite{Blumberger2015} compared to $\sim\!1 k_\mathrm{B}T$ for chlorophyll excitations. As a result of the large system-bath coupling strength, direct HEOM calculations involving CT states become very challenging.\cite{Shi2009,Shi2009b,Firmino2016} Recent developments using matrix product states,\cite{Shi2018,Borrelli2019,Ke2022} and their generalizations,\cite{Yan2021} or tree tensor networks\cite{Lindoy2019} to solve the HEOM can help alleviate this problem, but these methods are limited to linear coupling models between the charge transfer states and harmonic environments. 
Alternative numerically exact methods such as MACGIC-QUAPI\cite{Richter2019,Richter2019a} have been successfully applied to models of coupled EET and CT in reaction centers,\cite{Richter2019a} but dynamics with this method can still be difficult to converge for large system sizes and strong system-bath coupling. Approximate theories, namely modified-Redfield/generalized F\"orster theory \cite{Zhang1998,Scholes2000,Novoderezhkin2010}have been applied to study CT processes in light-harvesting complexes,\cite{Novoderezhkin2011,Novoderezhkin2016,Novoderezhkin2018} but these methods do not always accurately describe the EET dynamics in the absence of CT processes.\cite{Kreisbeck2014} Semi-classical approaches have also been used to study exciton dynamics,\cite{Kelly2011,Huo2011,Kim2012,Pfalzgraff2019,Runeson2020a,Mannouch2020,Runeson2022} but these methods often break down for systems with large system-bath couplings, as is encountered in CT processes. In order to facilitate the study of reaction center and CT quenching processes in photosynthesis an accurate and computationally efficient method that can describe coupled EET and CT processes is needed.

In this work we present a theory combining the HEOM method, which is used to model the EET dynamics of locally excited states, with quantum master equation approaches used to describe the charge transfer\cite{Sparpaglione1988,Fay2018,Fay2021,Fay2021c} and radiative processes.\cite{Kreisbeck2011} Conceptually similar hybrid approaches in which different degrees of freedom are treated with different levels of theory have been used previously to extend the applicability of approximate theories,\cite{thoss2001self,berkelbach2012reduced,berkelbach2012reduced,montoya2015extending,schile2019simulating} but here we take the numerically exact HEOM method and make it more computationally tractable through judicious approximations on a subset of dynamical degrees of freedom. The working equations of our method, obtained using a Zwanzig projection approach,\cite{Nakajima1958,Zwanzig1960,Mori1965} are a set of simple linear differential equations for hierarchies of auxiliary density operators for the different manifolds of states in the system. Formally the method can account for anharmonicity in the degrees of freedom coupled to the CT processes, though in this work we only consider linear response models for the CT process. 
 In Sec.~\ref{sec-theory} we outline the model and the HEOM method, and in Sec.~\ref{sec-hybridheom} we derive the hybrid HEOM/QME method. In Sec.~\ref{sec-dimer} we test the method against exact results for a dimer of locally excited states coupled to a CT state in order to verify the accuracy of the method. In Sec.~\ref{sec-lhcii} we then apply the hybrid HEOM/QME method to study charge transfer quenching in light-harvesting complex II (LHCII),\cite{Mirkovic2017,Muh2010,Novoderezhkin2010,Kreisbeck2014,Cupellini2020} a system which would be intractable to study with direct HEOM calculations. Our simulations of LHCII reveal the importance of the excitation energy funnel in determining photo-protective quenching efficiency in the LHCII complex, a result which we discuss further in Sec.~\ref{sec-disc}. Final we draw conclusions in Sec.~\ref{sec-conc}.

\section{Theory}\label{sec-theory}
We consider the coupled energy and charge transfer of a photoexcited chromophoric system, like that found in naturally occurring light harvesting complexes. In this section, we outline a general model hamiltonian, and review HEOM. 

\subsection{Exciton and charge transfer model}\label{sec-model}
\begin{figure}[t]
	\includegraphics[width=0.45\textwidth]{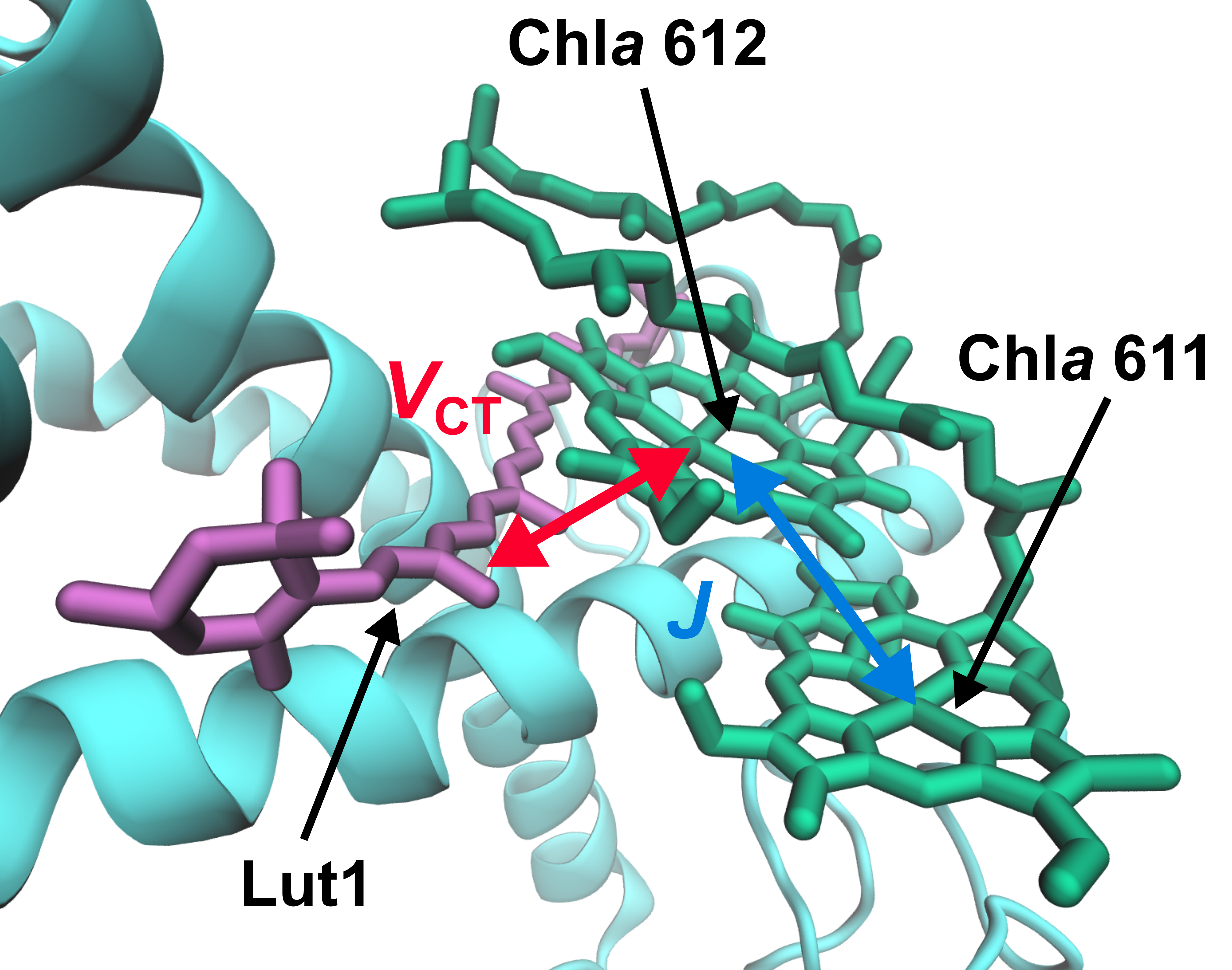}
	\caption{The structure of the \Chla 611-\Chla 612 dimer in LHCII with the lutein (Lut 620) electron donor (PDB 1RWT, chain C\cite{Liu2004}). The electron transfer coupling is denoted $V_{\CT}$ and the electrostatic inter-chromophore coupling is denoted $J$.}\label{chladimer-lut-fig}
\end{figure}
The system we consider consists of chromophores, and electron donors and acceptors. The chromophores have a ground state $\ket{\GS}$, and a manifold of singly excited states, which can be spanned by a local basis  $\ket{\LE_n}$, which are coupled electrostatically. Such a system of coupled LE states can be well described by a Frenkel exciton model.\cite{Mirkovic2017} These locally excited states can also undergo charge transfer, where either the excited electron or hole transfers to a nearby acceptor or donor. The charge transfer states that can be formed by these processes are denoted as $\ket{\CT_n}$. These states can undergo charge recombination to return the system to $\ket{\GS}$. An example of such a system is the \Chla\ 611--\Chla\ 612 dimer in LHCII shown in Fig.~\ref{chladimer-lut-fig},\cite{Cupellini2020} where the locally excited \Chla\ states couple to each other, and the locally excited \Chla\ 612 can also accept an electron from the nearby lutein donor. As well as coupling to each other, the LE and CT electronic excitations couple to the nuclear degrees of freedom on the chromophores, the donors/acceptors, and the surrounding polarizable environment, which leads to decoherence and relaxation of these excited electronic states. For example in the \Chla\ dimer in Fig.~\ref{chladimer-lut-fig} the local excitation on each \Chla\ couples primarily to the vibrations localized on each chlorophyll,\cite{Mirkovic2017} but the CT excitation couples to the the intramolecular \Chla\ and lutein vibrations, and the low frequency modes which determine polarization of the surrounding protein and solvent environment.\cite{Blumberger2015} Furthermore the electronic states of the system couple to the electromagnetic field, which creates radiative decay pathways for the excited electronic states.\cite{Kreisbeck2011} %\textit{\color{blue}Is this introduction of the model clear enough on its own? I was also thinking I could introduce it by specifically referencing the Chla-Lut CT quenching in LHCII, along with a picture to help ground this intro.}

The Hamiltonian for the system described above can be written as
\begin{align}
	\op{H} = \op{H}_{\LE} + \op{H}_{\CT} + \op{H}_\GS + \op{H}_{\LE,\CT} + \op{H}_{\CT,\GS} + \op{H}_\mathrm{EM} + \op{H}_{\mathrm{D}}
\end{align}
where the ground state Hamiltonian $\op{H}_\GS$ decomposes as
\begin{align}
	\op{H}_{\GS} =  \op{\Pi}_{\GS}(\op{T} + \op{V}_0), %\op{\Pi}_{\GS}\op{H}_0 =
\end{align}
here $\op{T}$ is the nuclear kinetic energy operator, $\op{V}_0$ is the ground-state potential energy operator, and $\op{\Pi}_{\GS}$ is a projection operator $\op{\Pi}_{\GS} = \dyad{\GS}$. Similarly the Hamiltonian of the charge transfer states is
\begin{align}
	\op{H}_\CT = \sum_{n=1}^{N_\CT} \op{\Pi}_{\CT_n} \op{H}_{\CT_n} = \sum_{n=1}^{N_\CT}\op{\Pi}_{\CT_n}(\op{T} + \op{V}_{\CT_n})
\end{align}
with $\op{\Pi}_{\CT_n} = \dyad{\CT_n}$ being a projection operator onto the $\CT_n$ state, which we assume is a CT state in which electrons and holes are localized on specific acceptors and donors. For the locally excited states we take a similar form but include all the LE state couplings $J_{nm}$,
\begin{align}
	\begin{split}
		\op{H}_{\LE} &= \bigg(\sum_{n=1}^{N_\LE} \dyad{\LE_n} (E_n + \op{T} + \op{V}_{\LE_n}) \\
		&+ \sum_{n>m}J_{nm}\left(\dyad{\LE_n}{\LE_m} + \dyad{\LE_m}{\LE_n}\right),
	\end{split}\\
&= \op{H}_{\LE,\sys} + \sum_{n=1}^{N_\LE} \dyad{\LE_n} (\op{T} + \op{V}_{\LE_n})
\end{align}
and we can again define an electronic projection operator $\op{\Pi}_{\LE} = \sum_n \dyad{\LE_n}$ which commutes with $\op{H}_{\LE}$. The CT-GS diabatic coupling term can be written as
\begin{align}
	\op{H}_{\CT,\GS} = \sum_{n=1}^{N_{\CT}} V_{\CT_n,\GS}\left(\dyad{\CT_n}{\GS} + \dyad{\GS}{\CT_n}\right)
\end{align}
and the locally excited state charge transfer diabatic coupling term can be written as 
\begin{align}
	\op{H}_{\LE,\CT} = \sum_{n=1}^{N_\LE}\sum_{m=1}^{N_\CT} V_{\LE_n,\CT_m}\left(\dyad{\LE_n}{\CT_m}+\dyad{\CT_m}{\LE_n} \right).
\end{align}
with coupling constants $V_{\CT_n,\GS}$ and $V_{\LE_n,\CT_m}$. For simplicity, we have made the Condon approximation, by assuming the diabatic state couplings have no nuclear coordinate dependence. 

The electromagnetic field Hamiltonian $\op{H}_\mathrm{EM}$ is given by\cite{Kreisbeck2011,Jankovic2020}
\begin{align}
	\op{H}_\mathrm{EM} = \sum_{\vb*{k},p}\hbar \omega_{\vb*{k}} \left(\op{a}^\dag_{\vb*{k}p}\op{a}_{\vb*{k}p}+\frac{1}{2}\right)
\end{align}
where $\op{a}_{\vb*{k}p}$ is the electromagnetic (EM) field annihilation operator for mode $\vb*{k}$ with polarization $p$, and $\omega_{\vb*{k}} = c_0 |\vb*{k}|$. These EM field modes are denoted the $\bath_{\EM}$ degrees of freedom. $\op{H}_\mathrm{D}$ is the dipole coupling operator between the molecular system and the EM field,\cite{Kreisbeck2011,Jankovic2020,Bennett2015}
\begin{align}
	\op{H}_\mathrm{D} = - \op{\vb*{\mu}}\cdot \op{\vb*{\mathcal{E}}}
\end{align}
within a point dipole approximation for the system. Here $\op{\vb*{\mu}}$ is the system transition dipole moment operator, with components for the $\LE_n$ state, $\op{\mu}_{n,\alpha}$, with $\alpha = x,y,z$, given by
\begin{align}
	\op{\mu}_{\alpha} = \sum_{n=1}^{N_{\LE}}\mu_{n,\alpha}\left(\dyad{\LE_n}{\GS} + \dyad{\GS}{\LE_n}\right),
\end{align}
and $\op{\vb*{\mathcal{E}}} $ is the electric field operator at the origin,
\begin{align}
	\op{\vb*{\mathcal{E}}} = i\sum_{\vb*{k},p} \sqrt{\frac{\hbar \omega_{\vb*{k}}}{2\pazocal{V}_0\varepsilon_0}}\left(\op{a}_{\vb*{k}p}\vb*{e}_{\vb*{k}p}-\op{a}_{\vb*{k}p}^\dag\vb*{e}_{\vb*{k}p}^*\right),
\end{align}
in which $\pazocal{V}_0$ is the volume of the system, $\varepsilon_0$ is vacuum permittivity, and $\vb*{e}_{\vb*{k}p}$ is a unit vector defining the polarization of the EM field mode $\vb*{k}p$.

In describing the potential energy surfaces for the different diabatic states, we start by separating out the nuclear bath coordinates that modulate the energy gap between the $\LE_n$ state and the ground-state, which we denote the ${\bath_{\LE_n}}$ degrees of freedom. 
This is justified by noting that the main degrees of freedom that modulate the $\LE_n$ energies are intramolecular vibrational modes on chromophore $n$ and its surrounding local environment. Given the $\CT_n$ states involve these chromophores, the $\CT_n$ energies can also depend on the ${\bath_{\LE_n}}$ degrees of freedom, as well as local modes on donors/acceptors involved in the charge transfer, and delocalized modes corresponding to environment polarization.\cite{Ishizaki2009a,Mirkovic2017} These additional bath degrees of freedom are denoted $\bath_{\CT}$. In what follows we assume that the $\bath_{\LE}$ and $\bath_{\CT}$ degrees of freedom are not coupled and are therefore uncorrelated, and operators on these degrees of freedom, indicated by superscript $\bath_{\LE}$ and $\bath_{\CT}$ labels, therefore commute. This assumption is analogous to separating intramolecular and environmental contributions to spectral densities used in modeling condensed phase optical spectra,\cite{Lee2016,Segarra-Marti2020,Zuehlsdorff2020} and the separation of inner and outer-sphere contributions to electron transfer reorganization energies.\cite{Blumberger2015,Firmino2016}

Within this assumption, we can write down a model Hamiltonian for the coupled LE and CT states as
\begin{align}
	\begin{split}
		\op{H} &= \op{H}_{\LE,\sys} + \sum_{n=1}^{N_{\LE}}\dyad{\LE_n} \Delta \op{V}_{\LE_n}^{\bath_{\LE_n}}\\
		&+\sum_{n=1}^{N_\CT} \op{\Pi}_{\CT_n}\left(\op{H}_{\CT_n,\sys} + \Delta \op{V}_{\CT_n}^{\bath_{\CT}} + \sum_{m=1}^{N_\LE}\Delta\op{V}_{\CT_n}^{\bath_{\LE_m}}\right)\\
		&+\op{H}_\GS^{\bath_{\LE}}   + \op{H}_\GS^{\bath_{\CT}}  +  \op{H}_{\mathrm{ET}} + \op{H}_\mathrm{EM} + \op{H}_{\mathrm{D}}.
	\end{split}
\end{align}
Here $\op{H}_{\LE,\sys}$ is the LE system Hamiltonian, containing the LE state energies and couplings, and similarly $\op{H}_{\CT_n,\sys}$ is the system Hamiltonian for $\CT_n$ which describes the energy of $\CT_n$, and $\op{H}_{\mathrm{ET}} =  \op{H}_{\LE,\CT} + \op{H}_{\CT,\GS}$. 
$\Delta \op{V}_{\LE_n}^{\bath_{\LE_n}} = \op{V}_{\LE_n} - \op{V}_0$ describes how the potential energy surface of the ground state is perturbed by the electronic excitation $\LE_n$, and similarly $\Delta \op{V}_{\CT_n}^{\bath_{\CT}}+\sum_m\Delta \op{V}_{\CT_n}^{\bath_{\LE_m}} = \op{V}_{\CT_n} - \op{V}_0$ describes how the ground state potential is shifted in the charge transfer state $\CT_n$. The potential energy shift for the CT states are divided into a sum of terms which are correlated with the $\LE_m$ energy shifts, $\Delta \op{V}_{\CT_n}^{\bath_{\LE_m}}$, corresponding to reorganization of the intramolecular modes on each chromophore, and the remaining uncorrelated component $\Delta \op{V}_{\CT_n}^{\bath_{\CT}}$. The reference ground state potentials for the $\bath_{\LE}$ and $\bath_{\CT}$ degrees of freedom are given by $\op{H}_\GS^{\bath_{\LE}}$ and $\op{H}_\GS^{\bath_{\CT}}$ respectively.

\subsection{The LE state potential energy surfaces}

We can often assume that the LE state energy shift operators $\Delta \op{V}_{\LE_n}^{\bath_{\LE_n}}$ have Gaussian statistics in the ground state reference ensemble, meaning third-order cumulants and higher vanish. Further we assume correlation functions of $\Delta \op{V}_{\LE_n}^{\bath_{\LE_n}}$ can be decomposed into a sum of contributions as follows\cite{Ishizaki2009,Ishizaki2009a}
\begin{align}
	\ev{\Delta {V}_{\LE_n}^{\bath_{\LE_n}}(t)\Delta \op{V}_{\LE_m}^{\bath_{\LE_m}}(0)}_{\bath_{\LE}} 
	&=\delta_{n,m} \sum_{r=1}^{N_{\bath,n}} C_{n,r}(t)
\end{align}
where $\ev{\cdots}_{\bath_{\LE}} = \Tr_{\bath_{\LE}}[\cdots e^{-\beta \op{H}_{\GS}^{\bath_{\LE}}}]/\Tr_{\bath_{\LE}}[e^{-\beta \op{H}_{\GS}^{\bath_{\LE}}}]$, and $\Delta {V}_{\LE_n}^{\bath_{\LE_n}}(t) = e^{i \op{H}_{\GS}^{\bath_{\LE}} t /\hbar} \Delta {V}_{\LE_n}^{\bath_{\LE_n}}e^{-i \op{H}_{\GS}^{\bath_{\LE}} t /\hbar}$. Provided  $C_{n,r}^{\bath_{\LE}}(t)$ is a smooth function it can be written in terms of a spectral density $\mathcal{J}_{n,r}(\omega) $ as
\begin{align}
C_{n,r}^{\bath_{\LE}}(t) = \frac{\hbar}{\pi}\!\!\int_0^\infty \!\!\!\!\dd{\omega} \mathcal{J}_{n,r}(\omega)\!\left[\coth(\frac{\beta\hbar\omega}{2})\cos(\omega t) \!-\! i\sin(\omega t) \right].
\end{align}
With these assumptions, $\op{H}_\GS^{\bath_{\LE}}$can be written as a sum of independent harmonic bath Hamitlonians, and that the LE state energy shift terms $\Delta \op{V}_{\LE_n}^{\bath_{\LE_n}}$ are linear in the bath mode displacements.\cite{Tanimura2020} These assumptions are widely used in describing exciton dynamics, and largely hold due to the relatively weak coupling between the LE states and the environment, meaning shifts in the potential energy surfaces can be well approximated as simple shifts in a harmonic potential.\cite{Ishizaki2009a,Mirkovic2017} Overall this means we can write the $\bath_{\LE}$ Hamiltonians as\cite{Ishizaki2009a}
\begin{align}
	\op{H}_{\GS}^{\bath_{\LE}} &= \sum_{n=1}^{N_\LE} \sum_{r=1}^{N_{\bath,n}} \sum_{\alpha=1}^{N_{n,r}}\left(\frac{\op{p}_{nr\alpha}^2}{2m_{nr\alpha}} + \frac{1}{2}m_{nr\alpha}\omega_{nr\alpha}^2\op{q}_{nr\alpha}^2\right) \label{Hb-eq}\\
	\Delta \op{V}_{\LE_n}^{\bath_{\LE_n}} &= \sum_{r=1}^{N_{\bath,n}} \op{B}_{n,r}=\sum_{r=1}^{N_{\bath,n}}\sum_{\alpha=1}^{N_{n,r}} c_{nr\alpha} \op{q}_{nr\alpha} \label{B-eq}
\end{align}
where the bath mode frequencies $\omega_{nr\alpha}$ and coupling coefficients $c_{nr\alpha}$ of the mode displacement operators $\op{q}_{nr\alpha}$ appear in the bath coupling operators $\op{B}_{nr\alpha}$. We can evaluate the spectral density
\begin{align}
	\mathcal{J}_{n,r}(\omega) = \frac{\pi}{2} \sum_{\alpha=1}^{N_{n,r}} \frac{c_{nr\alpha}^2}{m_{nr\alpha}\omega_{nr\alpha}}\delta(\omega - \omega_{nr\alpha})
\end{align} 
in terms of these microscopic parameters in the Hamiltonian.

% Neither \kappa_{m,r}^{\CT_n} not \op{B}_{m,r} are used in the rest of the manuscript
%We also assume that the $\Delta \op{V}_{\CT_n}^{\bath_{\LE_m}}$ terms, which describe the correlation between the CT state and the LE state energy shifts, can be written is linear in the $\bath_\LE$ bath coupling operators in Eq.~\eqref{B-eq}, with coupling coefficients $\kappa_{m,r}^{\CT_n}$,
%\begin{align}
%	\Delta \op{V}_{\CT_n}^{\bath_{\LE_m}} = \sum_{r=1}^{N_{m,r}} \kappa_{m,r}^{\CT_n}\op{B}_{m,r}.
%\end{align}
%With this description of how the LE state energies depend on the nuclear degrees of freedom, we can treat the dynamics of the LE states exactly with the hierarchical equations of motion method, which is described in the next section. 

\subsection{The hierarchical equations of motion}

The hierarchical equations of motion provide a method for simulating the dynamics of a system linearly coupled to a harmonic bath. It was developed for Hamiltonians like those that we are interested in that can be decomposed as\cite{Tanimura2020}
\begin{align}
	\op{H}  = \op{H}_\sys + \sum_{j=1}^{N_\bath}\left(\op{H}_{\bath,j} + \op{V}_j \op{B}_j \right).
\end{align}
where $\op{H}_\sys$ is the sub-system Hamiltonian, $\op{V}_j$ are system operators, $\op{H}_{\bath,j}$ is the Hamiltonian of harmonic bath $j$. The bath displacement operators, $\op{B}_j$, are defined as,
%\begin{align}
$\op{H}_{\bath,j} = \sum_{\alpha=1}^{N_{j}} {\op{p}_{j\alpha}^2/2m_{j\alpha}} + m_{j\alpha}\omega_{j\alpha}^2\op{q}_{j\alpha}^2/2$ %\\
and $\op{B}_{j} = \sum_{\alpha=1}^{N_{j}} c_{j,\alpha} \op{q}_{j,\alpha}.\label{Bj-eq}$ 
%\end{align}
as analogous to Eqs.~\eqref{Hb-eq} and \eqref{B-eq} where the index $n,r$ is replaced with a single index $j$. 
Assuming the bath displacement operator correlation functions can be decomposed as
\begin{align}
	C_j(t) &= \sum_{k=1}^{\infty} a_{jk} e^{-\nu_{jk} t} %\text{ and }	C_j^*(t)= \sum_{k=0}^{\infty} \bar{a}_{jk}^* e^{-\nu_{jk} t},
\end{align}
and using the Gaussian property of the harmonic baths, we can obtain the system reduced density operator $\op{\rho}(t) = \Tr_\bath[\op{\rho}_{\mathrm{tot}}(t)]$, from a hierarchy of auxiliary density operators (ADOs) $\op{\rho}_{\vb{n}}(t)$ which obey the following equation of motion,\cite{Tanimura1989,Ishizaki2005a,Tanimura2020}
\begin{align}
	\begin{split}
		\dv{t}\op{\rho}_{\vb{n}}(t) &= -\frac{i}{\hbar}\left[\op{H}_\sys, \op{\rho}_{\vb{n}}(t)\right] \!-\! \sum_{j,k} n_{jk}\nu_{jk}\op{\rho}_{\vb{n}}(t) \!+ \Xi_{\vb{n}}\op{\rho}_{\vb{n}}(t)  \\ 
		&-\frac{i}{\hbar}\sum_{j,k}\sqrt{(n_{jk}+1)|a_{jk}|}[\op{V}_j,\op{\rho}_{\vb{n}_{jk}^+}] \\
		&-\frac{i}{\hbar}\sum_{j,k}\sqrt{\frac{n_{jk}}{|a_{jk}|}}\left(a_{jk}\op{V}_j\op{\rho}_{\vb{n}_{jk}^-} -{a}_{jk}^*\op{\rho}_{\vb{n}_{jk}^-}\op{V}_j \right).
	\end{split}
\end{align}
where $\vb{n} = (n_{1,0}, n_{1,1},\dots,n_{j,k},\dots)$ is a multi-index that specifies the excitation level of mode $k$ for each bath $j$ for a given hierarchy element, $\vb{n}_{jk}^{\pm} =  (n_{1,0}, n_{1,1},\dots,n_{j,k}\pm 1,\dots)$, and $\Xi_{\vb{n}}$ is a system superoperator that accounts for finite truncation of the hierarchy. The sub-system reduced density operator can be obtained as the zeroth element of this hierarchy $\op{\rho}(t) = \op{\rho}_{\vb{0}}(t)$.

We can write down the truncated hierarchy as 
\begin{align}
	|\rho(t)\dra = \sum_{\vb{n}} |\rho_{\vb{n}} (t)\dra \otimes |\vb{n}\dra
\end{align}
where $|\vb{n}\dra$ is a basis vector corresponding to auxiliary density operator $\vb{n}$, and $|\rho_{\vb{n}} (t)\dra$ is the Liouville space vector of this ADO. With this notation we can write down the equation of motion more compactly as\cite{Shi2009b,Ikeda2022,Ke2022}
\begin{align}
	\dv{t}|\rho(t)\dra &= \pL |\rho(t)\dra \\
	&= (\pL_\sys\otimes \pazocal{I}_{\mathrm{ado}} - \pazocal{I}_\sys\otimes \Gamma + \Xi + \pazocal{V}) |\rho(t)\dra
\end{align}
where $\pL_\sys = -(i/\hbar)[\op{H}_\sys,\ \cdot\ ]$ is the system Liouvillian, $\pazocal{I}_\sys$ and $\pazocal{I}_{\mathrm{ado}}$ are identity operators on the system Liouville space and the set of ADOs respectively, $\Gamma = \sum_{\vb{n}}\gamma_{\vb{n}} |\vb{n}\dra\dla\vb{n}|$ is a matrix of ADO decay rates, $\Xi = \sum_{\vb{n} \in \mathcal{N}}\Xi_{\vb{n}} \otimes |\vb{n}\dra\dla\vb{n}|$ is superoperator that accounts for finite truncation of the hierarchy\cite{Ishizaki2005a}, and $\pazocal{V}$ is the term that couples different ADOs within the hierarchy. Henceforth we will swap between Liouville vector notation $|\rho\dra$ and standard operator notation $\op{\rho}$, as is most appropriate. 

% MOVE THIS DISCUSSION TO THE INTRO
%Although HEOM method has been widely used to study exciton dynamics of protein pigment complexes, directly using the HEOM to study charge transfer processes in these systems can be very difficult due to the large reorganization energies associated with charge transfer. Large reorganization energies correspond to large system-bath coupling strengths, which give rise to large values of $a_{jk}$ for baths associated with CT processes. The large coupling coefficients for certain modes means many ADOs are needed to converge the HEOM dynamics. This can make HEOM calculations involving CT states effectively impossible when there are large numbers of system states and baths. %Recent developments using matrix product states, the ML-MCTDH ansatz or tree tensor networks to solve the HEOM can help alleviate this problem, but these methods are still limited because they require they assume that the $\bath_{\CT}$ degrees of freedom can be treated as harmonic linearly coupled to the electronic states. 
%As reorganization energies become larger, non-linear coupling and anharmonic effect can contribute to electron transfer rates,\cite{Kuharski1988,Lawrence2020} thus it is desirable to find a method which can efficiently treat general $\bath_{\CT}$ degrees of freedom, whilst still treating the $\LE$ system-bath coupling with the HEOM approach. In the next section we will describe one such a method based on generalized quantum master equations.

\section{The hybrid HEOM/QME method}\label{sec-hybridheom}

%Dynamics within the LE state manifold cannot always be well-described by weak chromophore coupling or weak system-bath coupling theories, but instead dynamics often lie in an intermediate regime where system-bath coupling and inter-site couplings are comparable in strength.\cite{Ishizaki2010} This necessitates the use of a nonperturbative approach like HEOM to describe exciton dynamics. 
%Although HEOM method has been widely used to study exciton dynamics of protein pigment complexes, directly 
Using HEOM to study CT in these systems can be difficult due to the large reorganization energies, which give rise to large values of $a_{jk}$ for baths associated with CT processes. The large coupling coefficients for certain modes means many ADOs are needed to converge the HEOM dynamics. This can make HEOM calculations involving CT states intractable when there are large numbers of system states and baths.
%Additionally, as reorganization energies become larger, non-linear coupling and anharmonic effect can contribute to electron transfer rates,\cite{Kuharski1988,Lawrence2020} thus it is desirable to find a method which can efficiently treat general $\bath_{\CT}$ degrees of freedom, whilst still treating the $\LE$ system-bath coupling with the HEOM approach. 
%
A simplification results as the reorganization energy of charge transfer processes is typically much larger than the electronic coupling between CT states and LE states, the use of perturbation theory, where the coupling to the CT states is treated as a small parameter, can be expected to give an accurate description of the coupled dynamics of the LE and CT states.\cite{Blumberger2015,Fay2018,Fay2021} 
%Similarly coupling between LE states and the ground state via the dipole interaction with the electromagnetic field can be treated accurately with weak coupling perturbation theory between the system and the electromagnetic field.\cite{Kreisbeck2011} 
In this section, we describe how Zwanzig projection can be used to construct a hybrid HEOM/QME method in which radiative and electron transfer processes are treated with perturbation theory while preserving the high accuracy of the EET dynamics afforded by HEOM. Similar approaches have been used to derive hybrid HEOM methods to describe radiative processes in EET in protein-pigment complexes,\cite{Kreisbeck2011,Chan2018,Ko2021} but these have not included charge transfer processes. The approach taken here combines some of the ideas for describing radiative processes in Ref.~\onlinecite{Kreisbeck2011} with Nakajima-Zwanzig theory based approaches that have been used to derive quantum master equations for spin density operators of different charge transfer states in photo-excited molecules.\cite{Fay2018,Fay2021,Fay2021c}

\subsection{Zwanzig projection}

Here we use the Zwanzig projection operator approach to construct a hybrid HEOM/QME method. We set up the problem by formally including the $\bath_{\CT}$ and EM field degrees of freedom, as well as the full set of electronic states, in the system Hamiltonian $\op{H}_\sys$ in the hierarchical equations of motion. This means the ``system'' Hamiltonian is taken to be
\begin{align}
	\begin{split}
		\op{H}_\sys &= \op{H}_{\LE,\sys } + \sum_{n=1}^{N_{\CT}} \Pi_{\CT_n}\left(\op{H}_{\CT_n,\sys} + \Delta \op{V}_{\CT_n}^{\bath_{\CT}}\right) + \op{H}_{\GS}^{\bath_{\CT}} \\
		&+ \op{H}_{\LE,\CT} + \op{H}_{\CT,\GS} + \op{H}_\mathrm{EM} + \op{H}_{\mathrm{D}}.
	\end{split}	
\end{align}
The response of the system to the $\bath_{\LE}$ couplings is treated with the HEOM, with the set of $N_{\bath_{\LE}} = \sum_{n} N_{\bath,n} $ baths indexed by $j = n,r$, and 
\begin{align}
	\op{V}_{n,r} = \dyad{\LE_n} + \sum_{m=1}^{N_{\CT}} \kappa_{n,r}^{\CT_m} \op{\Pi}_{\CT_m} \, ,
\end{align}
are the set of system-bath coupling operators. The coefficients $\kappa_{n,r}^{\CT_m}$ here describe how the $\bath_{\LE}$ degrees of freedom are coupled to the CT states.

\begin{figure}[t]
	\includegraphics[width=0.4\textwidth]{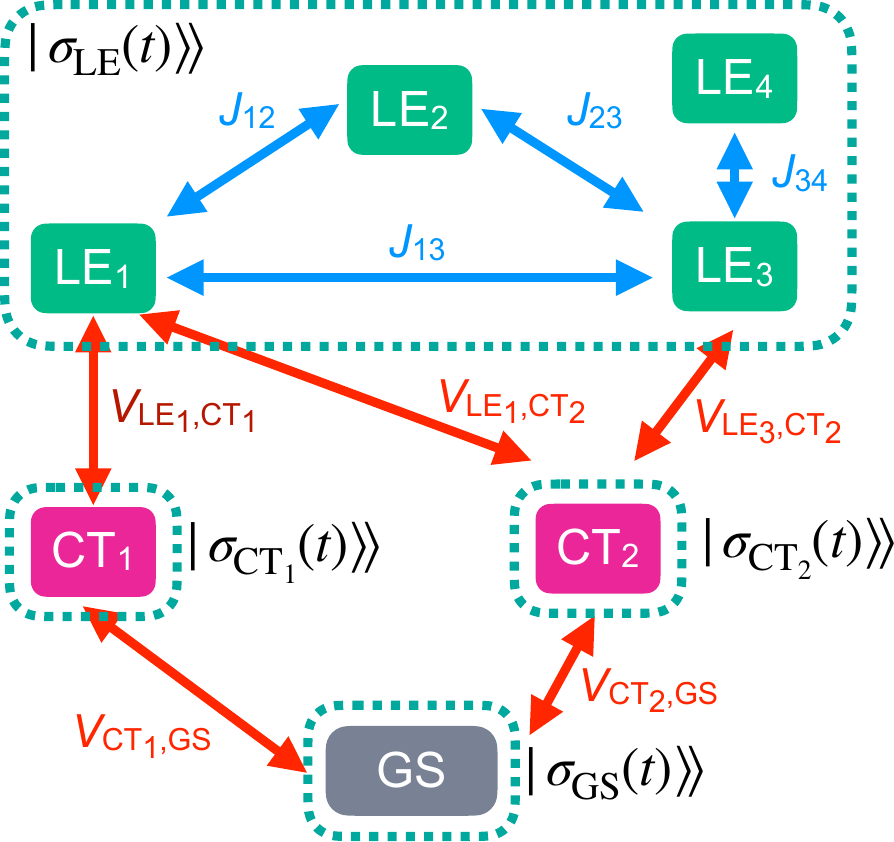}
	\caption{An schematic of the states, interstate coupling, and partitioning into reduced hierarchies $|\sigma_{A}(t)\dra$ for an example system with four LE states, two CT states and the GS.}\label{genscheme-fig}
\end{figure}

The main variables of interest for this system are the reduced density operator for the LE states, and the CT and GS state populations. These populations are given by
\begin{align}
	\op{\sigma}_{A,\sys}(t) = \op{\Pi}_A \Tr_{\bath}[\op{\rho}_{\vb{0}}(t)]\op{\Pi}_A
\end{align}
where $\op{\Pi}_A$ is a projection operator onto the manifold of electronic states with $A = \LE, \CT_n \text{ or } \GS$, and $\Tr_{\bath}$ denotes a trace over all $\bath_{\CT}$ and $\bath_{\EM}$ field degrees of freedom. Since we aim to treat the exciton dynamics with the HEOM, we construct the projection operator for the HEOM ADOs as
\begin{align}
	\pP &= \sum_{A}\pP_A \otimes \pazocal{I}_{\mathrm{ado}}\\
	\pP_A &= \op{\rho}^{\bath_{\CT}}_A \op{\rho}^{{\bath_{\EM}}} \op{\Pi}_A \Tr_\bath[ \ \cdot \ ]\op{\Pi}_A \, 
\end{align}
where local equilibrium density operator for the $\bath_{\CT}$ degrees of freedom in state $A$ is $\op{\rho}^{\bath_{\CT}}_A = e^{-\beta \op{H}_A^{\bath_{\CT}}}/\Tr_{\bath_{\CT}}[e^{-\beta \op{H}_A^{\bath_{\CT}}}]$, and $\op{H}_A^{\bath_{\CT}} = \op{H}_{\GS}^{\bath_{\CT}}$ for $A = \GS$ or $\LE$, and $\op{H}_{\CT_n}^{\bath_{\CT}} = \op{H}_{\GS}^{\bath_{\CT}} + \Delta \op{V}_{\CT_n}^{\bath_{\CT}}$. The EM field density operator is approximated as the zero temperature equilibrium density operator of the bare EM field, where all field modes are in their ground-state, $\op{\rho}^{{\bath_{\EM}}} = \bigotimes_{\vb*{k},p}\dyad{0_{\vb*{k}p}}$. This approximation is justified by the fact that we are only interested in spontaneous emission processes, since field modes at the LE state energies are not thermally excited at ambient temperatures, therefore $\beta \hbar \omega_{\vb*{k}p} \approx 0 $ for field modes resonant with the LE-GS energy gaps. This type of projection operator is analogous to that used in Refs.~\onlinecite{Fay2018} and \onlinecite{Fay2021} for electronic state spin density operators. This projected density operator contains the reduced density operator hierarchy, $|\sigma_{A}(t)\dra$, for each electronic state $A$,
\begin{align}
	\pP |\rho(t)\dra &= \sum_{A} |\rho_A^{\bath_{\CT}}\dra \otimes |\rho^{{\bath_{\EM}}}\dra \otimes |\sigma_{A}(t)\dra \\
	|\sigma_{A}(t)\dra &= \sum_{\vb{n}} |\sigma_{A,\vb{n}}(t)\dra \otimes |\vb{n}\dra \\
	\op{\sigma}_{A,\vb{n}}(t) &=  \op{\Pi}_A \Tr_{\bath}[\op{\rho}_{\vb{n}}(t)]\op{\Pi}_A.
\end{align}
An example of the partitioning into separate hierarchies for different state manifolds, $|\sigma_{A}(t)\dra$, is illustrated in Fig.~\ref{genscheme-fig} for system with four LE states, two CT states and the ground state.

Using this projection operator $\pP$ we can obtain a quantum master equation for the projected hierarchy of ADOs $\pP |\rho(t)\dra$ using the Nakajima-Zwanzig equation\cite{Nakajima1958,Zwanzig1960,Mori1965}
\begin{align}\label{nz-eq}
	\dv{t}\pP |\rho(t)\dra &= \pP \pL \pP |\rho(t)\dra + \!\!\int_0^t \!\!\pK(t-\tau)\pP |\rho(\tau)\dra \dd{\tau}\!\!,
\end{align}
where the memory kernel $\pK(t)$ is given by
\begin{align}\label{kernel-eq}
	\pK(t) = \pP \pL  e^{ (1-\pP)\pL t} (1-\pP)\pL \pP.
\end{align}
The generalized master equations for $|\sigma_{A}(t)\dra$ can be straightforwardly obtained from this by tracing out the $\bath_{\CT}$ and $\bath_{\EM}$ degrees of freedom. This gives the following general hybrid HEOM/QME for $|\sigma_{A}(t)\dra$,
\begin{align}\label{sigma-nz-eq}
	\dv{t}|\sigma_{A}(t)\dra \!&=\! \pL_A |\sigma_{A}(t)\dra + \!\sum_B\int_0^t \!\!\pK_{AB}(t-\tau) |\sigma_{B}(\tau)\dra\!\dd{\tau}\!.
\end{align}
where $\pL_A = \dla 1_\bath|\pP_A \pL \pP_A |\rho^\bath_A\dra$ is the component of $\pP \pL \pP$ that acts on $|\sigma_{A}(t)\dra$ and similarly $\pK_{AB}(t)$ is the component of $\pK_{AB}(t) = \dla 1_\bath|\pP_A \pK (t)\pP_B |\rho^\bath_B\dra $ that couples $|\sigma_{B}(t)\dra$ to $|\sigma_{A}(t)\dra$. Here $|1_\bath\dra$ is the identity operator on the $\bath_{\CT}$ and EM degrees of freedom and $|\rho_A^\bath\dra = |\rho_A^{\bath_{\CT}}\dra \otimes |\rho^{{\bath_{\EM}}}\dra$, and the Liouville space inner product is $\dla A | B \dra = \Tr [\op{A}^\dag \op{B}]$.

\subsection{Approximating the exact master equation}

While Eq.~\eqref{sigma-nz-eq} is formally exact,  evaluating the kernel requires evaluation of the exact dynamics generated by $\pL$, which due to the large Hilbert space associated with $\bath_{\CT}$ and $\bath_{\EM}$ is difficult. However, in the absence of the excitonic dynamics, each would be able to be treated accurately with simplifying approximations. Firstly, we make the Markovian approximation, in which we assume the decay time-scale of $\pK_{AB}(t)$ is much faster than the dynamics of $|\sigma_A(t)\dra$, and therefore we approximate the time-convolution terms as\cite{Sparpaglione1988,Fay2018}
\begin{align}
	\int_0^t \!\!\pK_{AB}(t-\tau) |\sigma_{B}(\tau)\dra\dd{\tau} \approx \int_0^\infty \!\!\pK_{AB}(\tau) \dd{\tau} |\sigma_{B}(t)\dra.
\end{align}
Secondly, we approximate the full kernel in Eq.~\eqref{kernel-eq} with its second order approximation in perturbation theory.\cite{Sparpaglione1988,Fay2018} Here the perturbation Liouvillian $\pL_V$ is taken to be
\begin{align}
	\pL_V &= \pL_{\mathrm{ET}} + \pL_{\mathrm{R}}\\
	\pL_{\mathrm{ET}} &= -\frac{i}{\hbar}[\op{H}_{\mathrm{ET}}, \ \cdot \ ]\otimes \pazocal{I}_{\mathrm{ado}} \\
	\pL_{\mathrm{R}} &= -\frac{i}{\hbar}[ \op{H}_{\mathrm{D}}, \ \cdot \ ]\otimes \pazocal{I}_{\mathrm{ado}},
\end{align}
and with this the kernel $\pK(t)$ can be approximated as
\begin{align}
	\pK(t) \approx \pK^{(2)}(t) = \pP \pL_V e^{(\pL_0+\pazocal{V})t} \pL_V \pP,
\end{align}
where we have defined $\pL_0 = \pL - \pL_V - \pazocal{V}$. Because the baths are uncorrelated, the second order kernel can be split into 
\begin{align}
	\pK^{(2)}(t) &= \pK^{(2)}_{\mathrm{ET}}(t) + \pK^{(2)}_{\mathrm{R}}(t)\\
	&=\pP \pL_\mathrm{ET} e^{(\pL_0+\pazocal{V})t} \pL_\mathrm{ET} \pP + \pP \pL_\mathrm{R} e^{(\pL_0+\pazocal{V})t} \pL_\mathrm{R} \pP
\end{align}
an electron transfer term, $\pK^{(2)}_{\mathrm{ET}}(t)$ and a radiative decay term, $\pK^{(2)}_{\mathrm{R}}(t)$.

The final simplification we make is to approximate the reference propagator $e^{(\pL_0+\pazocal{V})t}$ appearing in the ET kernel as
\begin{align}
	e^{(\pL_0+\pazocal{V})t} \approx e^{(\pL_0+\tilde{\pazocal{V}})t}
\end{align}
where $\tilde{\pazocal{V}} = \sum_{\vb{n},\vb{n'}\in \pazocal{N}_{\pK}}|\vb{n}\dra \! \dla \vb{n}| \pazocal{V}|\vb{n'}\dra \! \dla \vb{n'}|$ is $\pazocal{V}$ projected onto a subset of the ADOs. We make this approximation to reduce the density of the coupling between elements of $|\sigma_A(t)\dra$ in the equations of motion, and increase the computational efficiency of the method. 
%In practice we choose the subset of ADOs by using a looser cut-off parameter for the hierarchy than is used to generate the hierarchy of auxiliary density operators (details of this are explained below). 
For the radiative decay term we simply approximate $	e^{(\pL_0+\pazocal{V})t} \approx e^{\pL_{0,\sys}t}$, where $\pL_{0,\sys}$ is just the coherent system dynamics term appearing in $\pL_0$. In deriving explicit expressions for terms appearing in the hybrid HEOM/QME, it is important to note that $\pL_0$ and ${\pazocal{V}}$ can be decomposed as $\pL_0 = \bigoplus_{A,B} \pL_0^{AB}$ and ${\pazocal{V}} = \bigoplus_{A,B} \pazocal{V}^{AB}$, where each term only acts on coherences $\dyad{A}{B}$ between electronic states in manifolds $A$ and $B$.

Applying the above set of approximations to Eq.~\eqref{sigma-nz-eq}, allows us to write the hybrid HEOM/QME as
\begin{align}\label{heom-qme-eq}
	\dv{t}|\sigma_{A}(t)\dra \!&=\! \pL_A |\sigma_{A}(t)\dra +  \sum_{B}\pR_{AB}|\sigma_{B}(t)\dra ,
\end{align}
where $\pR_{AB}$ can be written as a sum of electron transfer and radiative decay terms, $\pR_{AB} = \pR_{AB}^{\mathrm{ET}} + \pR_{AB}^{\mathrm{R}}$. We can write this system of equations more explicitly as
\begin{align}
	\begin{split}
	&\dv{t}\op{\sigma}_{A,\vb{n}}(t) = -\frac{i}{\hbar}[\op{H}_{A,\sys},\op{\sigma}_{A,\vb{n}}(t)]  - \gamma_{\vb{n}} \op{\sigma}_{A,\vb{n}}(t) \\
	&+ \Xi_{A,\vb{n}}\op{\sigma}_{A,\vb{n}}(t) - \frac{i}{\hbar} \sum_{j,k} \sqrt{(n_{jk}+1)|a_{jk}|}[\op{V}_{A,j},\op{\sigma}_{A,\vb{n}_{jk}^+}(t)] \\
	&-  \frac{i}{\hbar} \sum_{j,k} \sqrt{\frac{n_{jk}}{|a_{jk}|}}\left(a_{jk}\op{V}_{A,j}\op{\sigma}_{A,\vb{n}_{jk}^-}(t)-{a}_{jk}^*\op{\sigma}_{A,\vb{n}_{jk}^-}(t)\op{V}_{A,j}\right)\\
	&+\sum_{B}\sum_{\vb{n}'} \pR_{AB,\vb{n}\vb{n}'} \op{\sigma}_{B,\vb{n}'}(t),
	\end{split}
\end{align}
where $\op{H}_{A,\sys}$ and $\op{V}_{A}$ are the components of the electronic state terms in $\op{H}_{\sys}$ and $\op{V}$ projected onto the manifold of electronic states $A$, $\Xi_{A,\vb{n}}$ is superoperator that accounts for finite truncation, projected on $A$, and $\pR_{AB,\vb{n}\vb{n}'} $ is the component of $\pR_{AB}$ that couples state $ \op{\sigma}_{A,\vb{n}}(t)$ to $ \op{\sigma}_{B,\vb{n}'}(t)$. Now that we have the general form of the Markovian hybrid HEOM/QME, we just need to evaluate the transfer operators $\pR_{AB}^{\mathrm{ET}}$ and $\pR_{AB}^{\mathrm{R}}$. This is detailed in the following sections. 

\subsection{The electron transfer term}

We start by considering the electron transfer term, $\pR_{AB}^{\mathrm{ET}}$. By noting that $\pL_0 + \tilde{\pazocal{V}}$ does not mix populations and coherences between the $\LE,\CT $ and $\GS$ manifolds, the elements of $\pR_{AB}^{\mathrm{ET}}$ where $A\neq B$ can be evaluated straightforwardly as
\begin{align}
	\pR_{AB}^{\mathrm{ET}} = \pL_{H_{\mathrm{ET}}}^{\mathrm{L},AB} \pG_{AB}^{B} \pL_{H_{\mathrm{ET}}}^{\mathrm{R},BA} +  \pL_{H_{\mathrm{ET}}}^{\mathrm{R},BA} \pG_{BA}^{B}  \pL_{H_{\mathrm{ET}}}^{\mathrm{L},AB}.
\end{align}
where the $\pL_{H_{\mathrm{ET}}}^{\mathrm{L/R},AB}$ terms are given by
\begin{align}
	\pL_{H_{\mathrm{ET}}}^{\mathrm{L},AB} \op{\sigma}&= \frac{1}{\hbar} \op{\Pi}_A\op{H}_{\mathrm{ET}}\op{\Pi}_B \op{\sigma} \\
\pL_{H_{\mathrm{ET}}}^{\mathrm{R},BA} \op{\sigma}&= \frac{1}{\hbar} \op{\sigma}\op{\Pi}_B\op{H}_{\mathrm{ET}}\op{\Pi}_A 
\end{align}
and the $\pG_{AB}^C$ terms are given by
\begin{align}
	\pG_{AB}^{B} = \int_0^\infty \dd{t} G_{AB}^{B}(t) \pazocal{S}_{AB}^{\mathrm{ET}} e^{\Lambda_{AB}^{\mathrm{ET}} t} {\pazocal{S}_{AB}^{\mathrm{ET}}}^{-1}
\end{align}
where we have used the spectral decomposition of the projected reference Liouvillian
\begin{align}
	(\pL_0^{AB} + \tilde{\pazocal{V}}_{\mathrm{ET}}^{AB}) \pazocal{S}_{AB}^{\mathrm{ET}} = \Lambda_{AB}^{\mathrm{ET}} \pazocal{S}_{AB}^{\mathrm{ET}} 
\end{align}
where $\pL_0^{AB}$ is defined as the block of $\pL_0$ that just acts on $AB$ coherences, and likewise for $\tilde{\pazocal{V}}_{\mathrm{ET}}^{AB}$, and $\pazocal{S}_{AB}^{\mathrm{ET}} $ is the matrix of eigenvectors and $\Lambda_{AB}^{\mathrm{ET}} $ is the diagonal matrix of eigenvalues. Due to the block diagonal structure of $\pL_0^{AB} + \tilde{\pazocal{V}}_{\mathrm{ET}}^{AB}$ this can be straightforwardly evaluated. The correlation function $G_{AB}^C(t)$ is given by
\begin{align}
	G_{AB}^C(t) &= \Tr_{\bath_{\CT}}[e^{-i\op{H}_A^{\bath_{\CT}}t/\hbar}e^{+i\op{H}_B^{\bath_{\CT}}t/\hbar}\op{\rho}_C^{\bath_{\CT}}]. 
\end{align}
which is the moment generating function for the energy gap between subsystems $A$ and $B$. 
The remaining diagonal $\pR_{AA}^{\mathrm{ET}}$ terms are given by
\begin{align}
	\pR_{AA}^{\mathrm{ET}} = -\sum_{B\neq A}\left(\pL_{H_{\mathrm{ET}}}^{\mathrm{L},BA} \pG_{BA}^{A} \pL_{H_{\mathrm{ET}}}^{\mathrm{L},BA} +  \pL_{H_{\mathrm{ET}}}^{\mathrm{R},AB} \pG_{AB}^{A}  \pL_{H_{\mathrm{ET}}}^{\mathrm{R},AB}\right) 
\end{align}
as required to conserve population.

%\textit{\color{blue}The rest of this section could be moved elsewhere, but I've put it here for now.} 
For practical calculations we use the full version of these expressions, but insight into the effect of ET on the exciton dynamics can be gained by making some additional approximations. First we consider the limit where the system and LE bath time-scales are long compared to the decay time of $G_{\CT,\LE}^\LE(t)$. In this limit the $\pR_{\LE,\LE}^{\mathrm{ET}}$ term becomes
\begin{align}
	\pR_{\LE,\LE}^{\mathrm{ET}} = -\sum_{n=1}^{N_\CT} \frac{k_{\CT_n\gets \LE}}{2}\left\{ \op{P}_{n},\ \cdot \ \right\} -\frac{i}{\hbar }\sum_{n=1}^{N_\CT} {\Delta_n}\left[\op{P}_{n},\ \cdot \ \right].
\end{align}
a sum of dissipative and conservative terms. Here, $k_{\CT_n\gets \LE}$ is the Fermi's Golden Rule rate constant for ET from the $\LE$ manifold to the $\CT_n$ state,\cite{Sparpaglione1988}
\begin{align}
	k_{\CT_n\gets \LE} &= \frac{|V_{\CT_n\gets \LE}|^2}{\hbar^2} \int_{-\infty}^\infty  G_{\CT,\LE}^{\LE}(t) e^{-i (E_{\CT_n}-\bar{E}_{\LE})t/\hbar}\dd{t} \label{fgr-site-eq}
\end{align}
and $\bar{E}_\LE$ is an effective energy scale of the LE states such that $\op{H}_{\LE,\sys} \approx \bar{E}_{\LE} \op{\Pi}_\LE$, $\op{H}_{\CT_n , \sys} = E_{\CT_n} \op{\Pi}_{\CT_n}$, with the effective coupling constant 
\begin{align}
	|V_{\CT_n\gets \LE}|^2 = \sum_{m=1}^{N_\LE} |V_{\LE_m,\CT_n}|^2
\end{align}
is given by a some over all of the $\LE$ states. Finally, $\op{P}_{n} = \dyad{\psi_{n}}$ is a projection operator onto the reactive state in the $\LE$ manifold which is given by
\begin{align}
	\ket{\psi_{n}} = \frac{1}{|V_{\CT_n\gets \LE}|}\sum_{m=1}^{N_{\LE}} \ket{\LE_m}V_{\LE_m,\CT_n},
\end{align}
and $\Delta_n$ is an energy shift term, 
\begin{align}
	\Delta_n = \frac{|V_{\CT_n\gets \LE}|^2}{\hbar} \Im\int_{0}^\infty  G_{\CT,\LE}^{\LE}(t) e^{-i (E_{\CT_n}-\bar{E}_{\LE})t/\hbar}\dd{t}
\end{align}
which in the limit of a highly activated reaction is given by
\begin{align}
	\Delta_n \approx \frac{	|V_{\CT_n\gets \LE}|^2}{\Delta E_{\CT_n\gets \LE}}
\end{align}
where $\Delta E_{\CT_n\gets \LE}$ is the vertical energy gap from the $\LE$ state to the $\CT_n$ state.\cite{Fay2018,Fay2021c}

From this we deduce that electron transfer has three main effects on the exciton dynamics. Firstly the ET causes population loss from the LE manifold from the reactive state $\ket{\psi_{n}}$. Secondly it causes decoherence between the states $\ket{\psi_{n}}$ state and the rest of the manifold of $\LE$ states. Thirdly, the energy shift terms perturb the exciton dynamics, modifying the energy gaps and couplings between LE states. If the reactive state is just a specific localized LE state, i.e. $\ket{\psi_{n}} = \ket{\LE_{r_n}}$, then the energy shift term just alters the energy of this $\LE$ state, which in the excitonic basis introduces coupling between delocalized excitonic states, as well as shifting exciton state energies. In the limit of Gaussian statistics for the bath and high temperature, this reduces to Marcus theory.\cite{Marcus1956,Hush1958,Sparpaglione1988} Eq.~\eqref{fgr-site-eq} ignores the effect of exciton formation, which shifts the free energy of states in the LE manifold, thereby changing the rate constants for electron transfer. The full hybrid HEOM/QME theory that we use in simulations however includes this important effect.

\subsection{The radiative decay term}

The radiative coupling terms can be derived in a similar manner to the ET term, details of which are given in the appendix. Here we simply state the final expressions for the two non-zero radiative transfer terms $\pR_{\GS,\LE}^{\mathrm{R}}$ and $\pR_{\LE,\LE}^{\mathrm{R}}$. The $\GS \gets \LE$ transfer term is given by
\begin{align}\label{R-rad-GSLE-eq}
	\begin{split}
		\pR_{\GS,\LE}^{\mathrm{R}} &= {\frac{ 1}{6 \hbar \varepsilon_0 c_0^3 \pi}}\!\!\sum_{\alpha=x,y,z}\!\! \bigg( \pL_{\alpha}^{\mathrm{L}} \pazocal{S}_{\LE,\GS}^{0,\sys} \Omega_{\LE,\GS}^3 ({\pazocal{S}_{\LE,\GS}^{0,\sys}})^{-1} \pL_{\alpha}^{\mathrm{R}} \\
		&+\pL_{\alpha}^{\mathrm{R}} \pazocal{S}_{\GS,\LE}^{0,\sys} \Omega_{\GS,\LE}^3({\pazocal{S}_{\GS,\LE}^{0,\sys}})^{-1} \pL_{\alpha}^{\mathrm{L}}  \bigg).
	\end{split}
\end{align}
and similarly the $\pR_{\LE,\LE}^{\mathrm{R}}$ term can be evaluated as
\begin{align}\label{R-rad-LELE-eq}
	\begin{split}
		\pR_{\LE,\LE}^{\mathrm{R}} &= -{\frac{ 1}{6 \hbar \varepsilon_0 c_0^3 \pi}}\!\!\sum_{\alpha=x,y,z}\!\! \bigg( \pL_{\alpha}^{\mathrm{R}} \pazocal{S}_{\LE,\GS}^{0,\sys} \Omega_{\LE,\GS}^3 ({\pazocal{S}_{\LE,\GS}^{0,\sys}})^{-1} \pL_{\alpha}^{\mathrm{R}} \\
		&+\pL_{\alpha}^{\mathrm{L}} \pazocal{S}_{\GS,\LE}^{0,\sys} \Omega_{\GS,\LE}^3({\pazocal{S}_{\GS,\LE}^{0,\sys}})^{-1} \pL_{\alpha}^{\mathrm{L}}  \bigg).
	\end{split}
\end{align}
where $\pL^\mathrm{L}_\alpha\op{\sigma} = \op{\Pi}_\GS \op{\mu}_\alpha \op{\Pi}_\LE \op{\sigma}$ and $\pL^\mathrm{R}_\alpha\op{\sigma} =  \op{\sigma}\op{\Pi}_\LE \op{\mu}_\alpha \op{\Pi}_\GS$. These expressions use the eigenvalue decompositions of the blocks of $\pL_{0,\sys}$, $\pL_{0,\sys}^{\GS,\LE}  = \pazocal{S}_{\GS,\LE}^{0,\sys} (i\Omega_{\GS,\LE}) ({\pazocal{S}_{\GS,\LE}^{0,\sys}})^{-1}$ and $\pL_{0,\sys}^{\LE,\GS}  = \pazocal{S}_{\LE,\GS}^{0,\sys} (-i\Omega_{\LE,\GS}) ({\pazocal{S}_{\LE,\GS}^{0,\sys}})^{-1}$, where $\Omega_{\GS,\LE}$ and $\Omega_{\LE,\GS}$ are diagonal matrices with real positive-valued entries. It should be noted that the decay rates for the excitonic states $\ket{\epsilon_n}$ appearing in these expressions exactly correspond to the standard Wigner-Weisskopf decay rates.\cite{Chan2018}

%\section{Results}

%Having outlined the theory of how to combine perturbation theory with the HEOM to treat electron transfer and radiative decay processes in excitonic systems, we shall now present two applications of this theory. Firstly, to validate the HEOM/QME method, we simulate the coupled exciton-charge transfer dynamics in a model system consisting of a dimer for LE states coupled to a single CT state. After this we apply the method to study Chlorophyll-Lutein charge transfer quenching in LHCII.

%\begin{figure}[b]
%	\includegraphics[width=0.35\textwidth]{fig3-dimerscheme}
%	\caption{A schematic diagrams showing the LE states and CT state in the dimer model, and the non-zero interstate couplings, together with the partitioning into reduced hierarchies.}\label{dimerscheme-fig}
%\end{figure}
\section{Exciton dimer model}\label{sec-dimer}

\begin{figure*}[t]
	\includegraphics[width=0.95\textwidth]{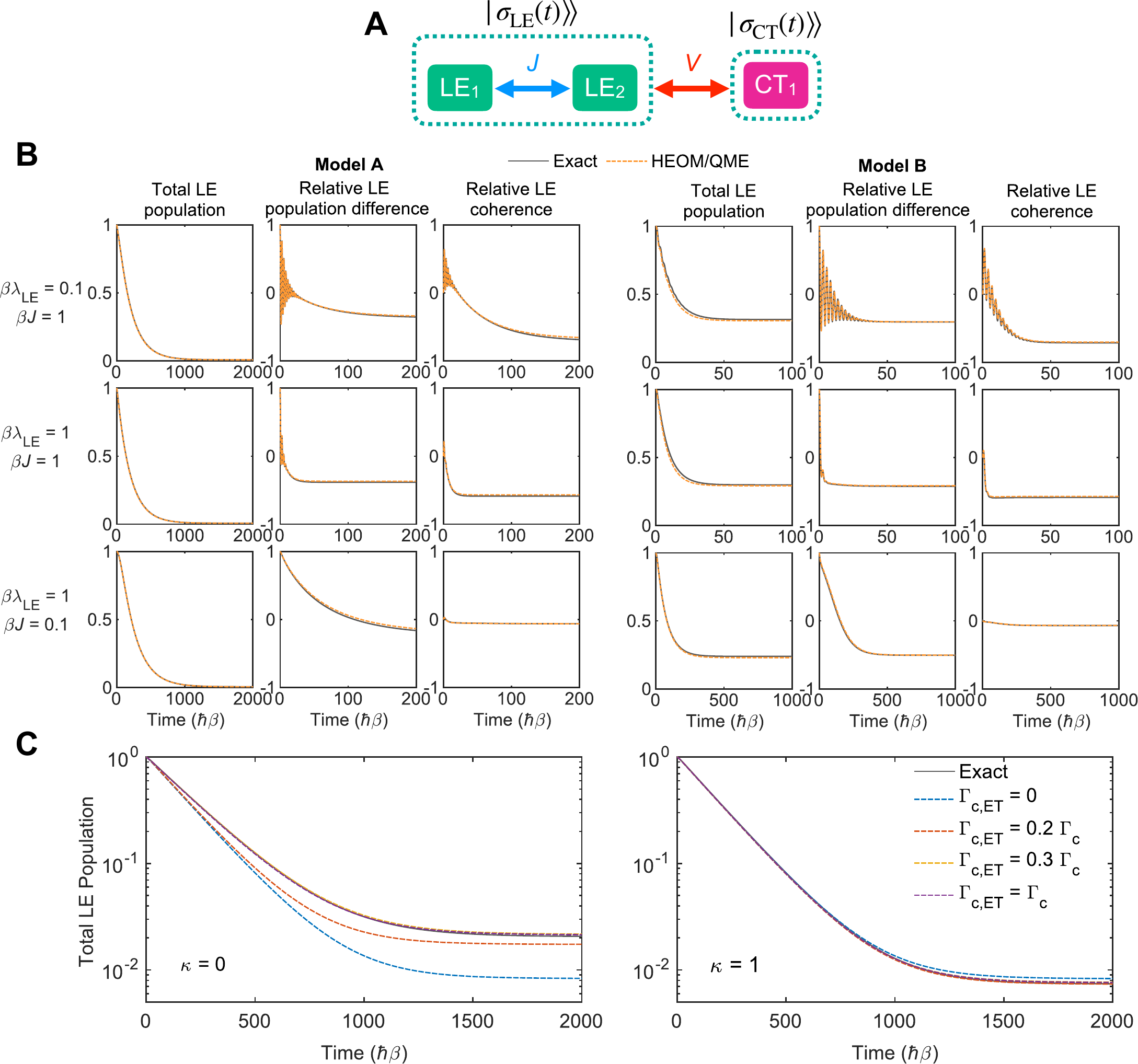}
	\caption{A: Schematic diagram showing the LE states and CT state in the dimer model, and the non-zero interstate couplings, together with the partitioning into reduced hierarchies. B: A comparison of the hybrid HEOM/QME method with exact HEOM result for the dimer model A with $\beta\hbar \gamma_{\mathrm{D}} = 0.25$, $\beta \Delta\epsilon = 1$, $\beta V = 0.1$, $\beta\lambda_{\CT} = 5$, $\kappa = 1$, and $\beta\Delta E_\CT = -6 $ (left three columns) and for the dimer model B with $\beta\hbar \gamma_{\mathrm{D}} = 1.75$, $\beta \Delta\epsilon = 1$, $\beta V = 0.5$, $\beta\lambda_{\CT} = 5$, $\kappa = 1$, and $\beta\Delta E_\CT = -2 $ (right three columns), with a range of values of $\lambda_{\LE}$ and $J$. The first/fourth columns show the total LE population dynamics, the second/fifth columns show the LE population difference relative to the total LE population, and the third/sixth columns show the real part of the LE coherence relative to the total LE population. The rows correspond to models with $\beta J = 1, \beta \lambda_{\LE} = 0.1$ (top), $\beta J = 1, \beta \lambda_{\LE} = 1$ (middle), and $\beta J = 0.1, \beta \lambda_{\LE} = 1$ (bottom). C: Convergence of the hybrid HEOM/QME results with respect to $\Gamma_{\mathrm{c, ET}}$ with parameters described in the text. The panels correspond to $\kappa = 0$ (left) and $\kappa = 1$ (right).}\label{dimer1-fig}\label{dimer3-fig}
\end{figure*}

%\subsection{Model details}
As a test for the approximations that go into the hybrid HEOM/QME method, we have performed simulations on a model exciton dimer, consisting of two LE states, coupled to a single CT state. The $\bath_{\CT}$ bath is taken to be harmonic in this example, allowing us to obtain exact dynamics directly with the HEOM method. The LE system Hamiltonian for this model is
\begin{align}
	\begin{split}
		\op{H}_{\LE,\sys} &= \frac{\Delta\epsilon}{2}\dyad{\LE_1} - \frac{\Delta\epsilon}{2}\dyad{\LE_2} \\
		&+ J(\dyad{\LE_1}{\LE_2} + \dyad{\LE_2}{\LE_1})
	\end{split}
\end{align}
and the CT state Hamiltonian is taken to be $\op{H}_{\CT,\sys} = (\Delta E_\CT + \lambda_{\CT})\dyad{\CT}$ and the CT bath shift operator is taken as $\Delta \op{V}_{\CT}^{\bath_{\CT}} =\op{B}_{\CT} $, where $\op{B}_{\CT}$ is a harmonic bath displacement operator. The LE baths and the $\bath_{\CT}$ bath, are taken to have Debye spectral densities,\cite{Tanimura1989}
\begin{align}
	\pazocal{J}_j(\omega) = \frac{\lambda_j}{2}\frac{\gamma_\mathrm{D} \omega}{\gamma_\mathrm{D}^2+\omega^2}
\end{align}
where $j = \LE_1,\LE_2$ or $\CT$ labels the bath. In this model we set $\lambda_{\LE_1} = \lambda_{\LE_2} = \lambda_{\LE}$ and only the $\LE_2$ state is coupled to the $\CT$ state with a coupling coefficient $V_{\LE_2,\CT} = V$. The correlation coefficients $\kappa_{\LE_n}^{\CT}$ describing the correlation between the CT state energy gap and the LE state energy gaps are taken to be $\kappa_{\LE_1}^{\CT} = 0$ and $\kappa_{\LE_2}^{\CT} = \kappa$. The structure of the coupling between states and the partitioning into different HEOM/QME hierarchies, $|\sigma_{A}(t)\dra$, is illustrated in Fig.~\ref{dimer1-fig} A.

The HEOM/QME calculations were performed with the adaptive short iterative Arnoldi integrator described in the SI of {Ref.~\onlinecite{fay2022simple}}, with a Krylov subspace dimension of $k=9$, and an error tolerance parameter of $\epsilon = 10^{-12}$. The hierarchy was truncated using the frequency cut-off criterion, with ADOs with $\gamma_{\vb{n}} > \Gamma_\mathrm{c} = 10 \gamma_{\mathrm{D}}$ excluded from the hierarchy. The same cut-off scheme was used to truncate $\tilde{\pazocal{V}}$ in evaluating the electron transfer kernel, but with a looser choice of cut-off parameter $\Gamma_\mathrm{c,ET} = 2.5 \gamma_{\mathrm{D}}$, so overall couplings between only 6 of the 55 ADOs were accounted for in evaluating $\pR_{AB}^{\mathrm{ET}}$. The HEOM was closed using the termination scheme and low temperature correction described in Ref.~\onlinecite{fay2022simple}.

%\begin{figure*}[t]
%	\includegraphics[width=0.9\textwidth]{fig5-dimer2.pdf}
%	\caption{Same as Fig.~\ref{dimer1-fig} but for models with $\beta\hbar \gamma_{\mathrm{D}} = 1.75$, $\beta \Delta\epsilon = 1$, $\beta V = 0.5$, $\beta\lambda_{\CT} = 5$, $\kappa = 1$, and $\beta\Delta E_\CT = -2 $.}\label{dimer2-fig}
%\end{figure*}

The exact HEOM calculations on the dimer model were also performed using the adaptive short iterative Arnoldi integrator with $k=9$ and $\epsilon = 10^{-11}$. The problem was simplified by reducing the number of baths from three to two as described in Appendix \ref{app-dimerbaths} -- this greatly reduced the number of ADOs needed in the exact calculations. The hierarchy was truncated using a reorganization energy weighted frequency cut-off scheme, wherein ADOs with a weight $w_{\vb{n}} = \sum_{jk}\nu_{jk}n_{jk}/\lambda_j > \tilde{L}_\mathrm{c}$ are excluded from the hierarchy, with $\tilde{L}_\mathrm{c} = 20$. This scheme accounts for the fact that the hierarchy needs to be deeper for the modes with of larger reorganization energy baths, because coupling coefficients down the hierarchy scale as $a_{jk} \propto \sqrt{\lambda_j}$. This more efficiently truncates the HEOM than other schemes, such as the frequency cut-off scheme\cite{Dijkstra2012} or the $L$, $M$ cut-off scheme.\cite{Ishizaki2005a}

%\subsection{Results}
As a first example of the HEOM/QME method, we performed simulations for the dimer model with $\beta\hbar \gamma_{\mathrm{D}} = 0.25$, $\beta \Delta\epsilon = 1$, $\beta V = 0.1$, $\beta\lambda_{\CT} = 5$, $\kappa = 1$, and $\beta\Delta E_\CT = -6 $ (labelled model A in Fig.~\ref{dimer1-fig} B), with a range of values of $\lambda_{\LE}$ and $J$. The initial condition was set to $\op{\sigma}_{\LE}(0) = \dyad{\LE_1}$ and $\op{\sigma}_{\CT}(0) = 0$. We look at three regimes of the exciton dynamics: the damped coherent transport regime ($\beta \lambda_{\LE} = 0.1$ and $\beta J = 1$), the vibrationally assisted transport regime ($\beta \lambda_{\LE} = 1$ and $\beta J = 1$), and the incoherent F\"orster transport regime ($\beta \lambda_{\LE} = 0.1$ and $\beta J = 0.1$). These model parameters are chosen to be typical of coupled LE and CT states in light harvesting complexes such as the \Chla\ dimer coupled to lutein in Fig.~\ref{chladimer-lut-fig}.\cite{Kreisbeck2014}

In Fig.~\ref{dimer1-fig} B we compare the exact simulated dynamics to the hybrid HEOM/QME dynamics, looking at three observables: the total LE state population, $p_{\LE}(t)$, the relative population difference between the LE states, $(p_{\LE_1}(t) - p_{\LE_2}(t))/p_{\LE}(t)$, and the relative LE coherence, defined as $2\Re[\mel{\LE_1}{\op{\sigma}_{\LE,\sys}(t)}{\LE_2}]/p_{\LE}(t)$. We see that in this example the hybrid HEOM/QME method performs excellently in the three regimes of exciton dynamics, with only small deviations in the long time limits of the relative LE population differences and coherences. In particular the time scale of decay of the LE population is captured very well by the hybrid method for all three models. 
%\begin{figure}[b]
%	\includegraphics[width=0.45\textwidth]{fig6-kappacompdimer.pdf}
%	\caption{Convergence of the hybrid HEOM/QME results with respect to $\Gamma_{\mathrm{c, ET}}$ with parameters described in the text. The panels correspond to $\kappa = 0$ (top) and $\kappa = 1$ (bottom).}\label{dimer3-fig}
%\end{figure}

In a second more challenging test for the hybrid method, we performed simulations for the same dimer model with $\beta\hbar \gamma_{\mathrm{D}} = 1.75$, $\beta \Delta\epsilon = 1$, $\beta V = 0.5$, $\beta\lambda_{\CT} = 5$, $\kappa = 1$, and $\beta\Delta E_\CT = -2 $ (labelled model B in Fig.~\ref{dimer1-fig} B). In this example the coupling between the CT and the LE manifold is much larger, the free energy of CT state is higher, so back reaction effects are more significant. Furthermore the characteristic bath frequency is comparable now to the LE system frequencies, leading to significant non-Markovian effects in the LE dynamics, and the ET rate from the LE manifold to the CT state is closer to frequencies of the exciton dynamics. However in Fig.~\ref{dimer1-fig} A we see that the hybrid method still performs well in all regimes of exciton dynamics, with only small errors in the decay of the total LE population. The errors in the population dynamics can likely be attributed to the increased importance of CT-LE coherence in this example, which is treated with a perturbative-Markovian approximation with the hybrid method. Within high temperature perturbation theory, the steady-state average CT-LE coherence scales roughly as $V/\lambda_{\CT} = 0.1$, which is five times larger in this set of models compared to the previous ones with $V/\lambda_{\CT} = 0.02$, so it is not unexpected that the rate of ET is not captured quite as well in these more challenging models.

As a final example, we examine the convergence of the LE population dynamics with respect to the cut-off parameter $\Gamma_\mathrm{c,ET}$, for two values of $\kappa=0$ and $\kappa = 1$. When $\kappa = 1$, there is no contribution from the $\bath_{\LE_2}$ bath to the free energy change or reorganization energy of the charge transfer process, however when $\kappa = 0$ the $\bath_{\LE_2}$ reorganizes in the $\LE_2\to \CT$ transfer, which means when $\kappa = 0 $ this bath has a significant contribution to the free energy change and reorganization energy of the $\LE_2\to \CT$ process. In these simulations, we set $\Gamma_\mathrm{c} = 7 \gamma_{\mathrm{D}}$ and we examine the population dynamics for the $\beta \lambda_{\LE} = 1$, $\beta J = 1$ with model parameters the same as in dimer model A (as in Fig.~\ref{dimer1-fig} B). In Fig.~\ref{dimer3-fig} C we show the population dynamics for various $\Gamma_\mathrm{c,ET}$ values for $\kappa = 0$ and $\kappa =1 $. We see that convergence is much faster in the $\kappa = 1$ case, where the LE bath does not contribute to the reorganization energy or free energy change of the LE$\to$CT process, so only the LE system energies need to be accounted for in the kernel to obtain accurate results. Conversely, when $\kappa = 0$ there is a significant contribution to the reorganization energy and free energy change from the LE bath (roughly 20\%) and therefore the LE bath response has to be accounted for in more detail in the ET kernel to obtain accurate results. 

\begin{figure*}[t]
	\includegraphics[width=0.95\textwidth]{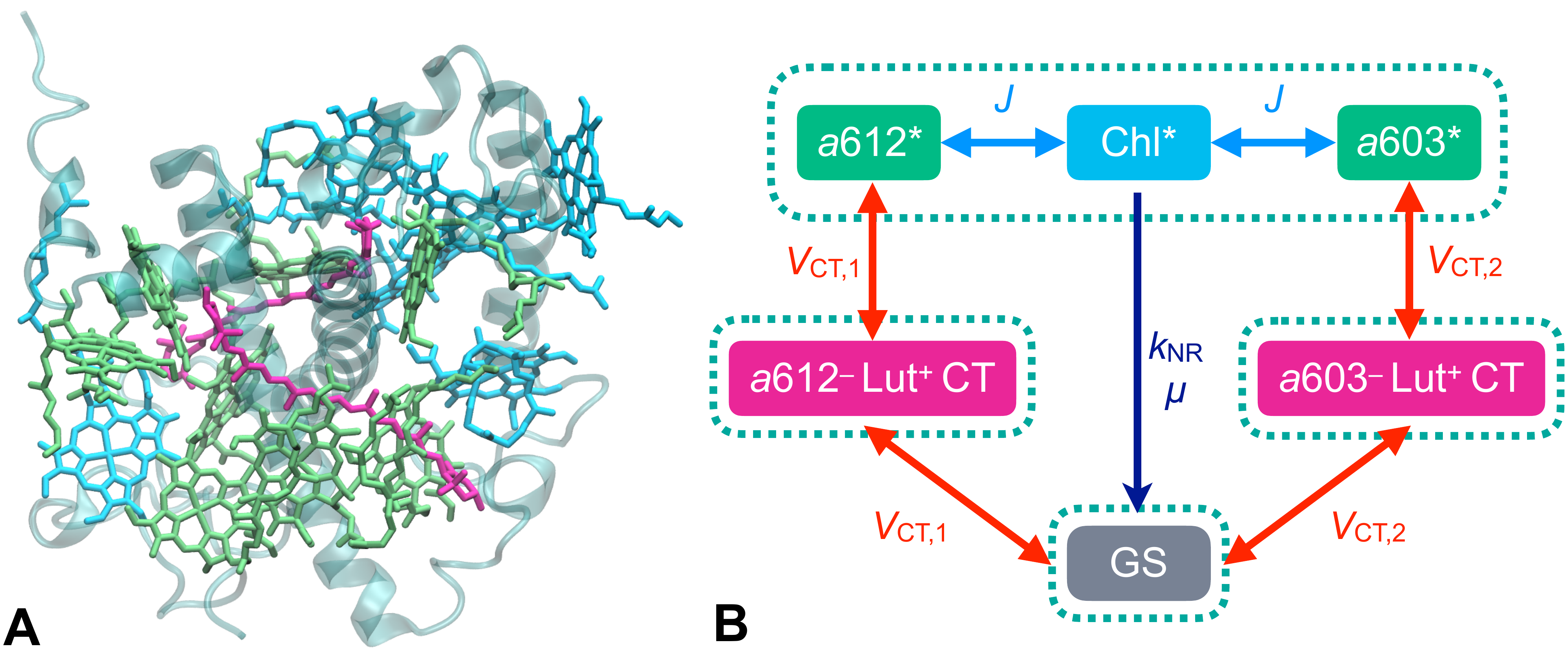}
	\caption{A: The structure of the LHCII monomer (PDB 1RWT chain C\cite{Liu2004}) showing the chlorophyll A (green), chlorophyll B (light blue), and lutein (pink) molecules. B: A scheme of the electronic states and couplings used in this model. Couplings between states are represented by double-headed arrows, the single-headed arrow represents the radiative and non-radiative transfer processes, and the dashed lines represent the blocks of states treated with their own explicit hierarchies of ADOs $|\sigma_A(t)\dra$.}\label{lhcii-model-fig}
\end{figure*}

\section{LHCII}\label{sec-lhcii}
Having established the accuracy of the hybrid HEOM/QME method on a range of dimer models, we turn to a more complex problem, charge transfer energy quenching in LHCII. LHCII is an important light-harvesting complex in plants, which absorbs light energy and transports it to reaction centers.\cite{Mirkovic2017} It is also known to play a role in non-photochemical quenching in plants, and one mechanism for excitation energy quenching in the comple x is electron transfer from the carotenoid lutein to excited chlorophyll-\textit{a} molecules.\cite{Goss2015,Ruban2021,Ruban2022,Cupellini2020} The resulting \ce{Chl\textit{a}^{$\bullet-$}Lut^{$\bullet+$}} pairs recombines to the ground state of the system, thereby quenching excitation energy as heat. In Ref.~\onlinecite{Cupellini2020} Cupellini \textit{et al.} parametrized the free energy change, reorganization energy and diabatic coupling for charge transfer from Lut1 to \chla{612}* and Lut2 to \chla{603}*, and from this they used Marcus theory and a simple kinetic model to estimate the excitation lifetime of chlorophyll in LHCII. This model treated the coupled \Chla* and \Chlb* dynamics with a simple kinetic model, where EET between the \chla{612}* and \chla{603}* states and the pool of \Chla* states is modelled as a simple first order rate process. This treatment ignores many of the details of EET in LHCII, such as the strong coupling between the \chla{612}* and \chla{611}* states, and the \chla{603}* and \chla{602}* states, which leads to exciton formation. 

\subsection{Model details}

In order to go beyond a simple kinetic treatment of excitation energy transfer and CT quenching in LHCII, we have modeled the coupled exciton and charge transfer dynamics of an LHCII monomer with the hybrid HEOM/QME method. This allows us to fully explore the effects of coupled exciton and charge transfer dynamics on the Chl* lifetime. The monomer contains eight \Chla\ and six \Chlb\ molecules, excitations on which are all coupled, together with two lutein molecules (Lut1 and Lut2), as shown in Fig.~\ref{lhcii-model-fig} A. We partition the system into an LE space, two CT states and the GS as outlined in Fig.~\ref{lhcii-model-fig} B. We explore this system using two different LE Hamiltonians for LHCII, which have been parametrized using various spectroscopic data: the model of M\"uh \etal\ from Ref.~\onlinecite{Muh2010} (henceforth referred to as model 1) and the model of Novoderezhkin \etal\ from Ref.~\onlinecite{Novoderezhkin2011} (henceforth referred to as model 2). These models differ subtly in the \Chla* and \Chlb* site energies and couplings, which is shown below to have a significant effect on the quenching dynamics. The LE-environment coupling is treated with the HEOM approach, with a single Debye bath for each site, with a reorganization energy of 220 cm${}^{-1}$ and characteristic frequency, $\gamma_{\mathrm{D},\ce{Chl}^*}$, of 353.7 cm${}^{-1}$, as taken from Kreisbeck \etal's study in Ref.~\onlinecite{Kreisbeck2014}. This model for the LE-environment coupling misses some small vibrational resonance effects in the \Chlb* to \Chla* energy transfer dynamics, but it accurately captures the important features of the excitation energy dynamics when compared with more complex structured environment models.\cite{Kreisbeck2014}

In order to describe the charge transfer process we use the reorganization energies, charge transfer free energy changes, and diabatic couplings calculated by Cupellini \etal\ from QM/MM simulations. The reorganization energy for the $\bath_{\CT}$ bath for each charge recombination process to the GS is assumed to the be same as for the corresponding charge separation, which is a reasonable assumption if polarization of the environment and lutein reorganization are the dominant contributions to the total reorganization energy, and the free energy of the GS relative to the excited states is taken from the LE model Hamiltonians. Using these parameters, we assume the $\bath_{\CT}$ bath can be treated as harmonic, with a Debye spectral spectral density with $\gamma_{\mathrm{D}} = 30\text{ cm}^{-1}$, which is representative of the response of the polarizable environment.\cite{Bennett2013} We have also explored how adding structure to this spectral density affects the quenching dynamics, as will be explained below. For the Lut1 $\to$ \chla{612}* charge transfer and Lut2 $\to$ \chla{603}* charge transfer we used the same diabatic coupling matrix elements as calculated by Cupellini \etal,\cite{Cupellini2020}  denoted $V_{\CT 1}$ and $V_{\CT 2}$ respectively. We use the same diabatic couplings for the recombination processes \chla{612}${}^-$ $\to$ Lut1${}^+$ and \chla{603}${}^-$ $\to$ Lut2${}^+$, given that the donor-acceptor separation is the same for the charge separation and charge recombination steps, although a different \Chla\ orbital is involved in the recombination process, so the couplings will be different in reality. The full set of parameters and the LE state Hamiltonians are given in Appendix \ref{lhcii-model-app}.

The transition dipole moment operators for the LE system are calculated using the atomic positions from PDB 1RWT\cite{Liu2004} for the \Chla\ and \Chlb\ molecules and assuming the the transition dipole moment for each LE state, $\vb*{\mu}_{n}$, points in the direction from \ce{N_B} to \ce{N_D}.\cite{BourneWorster2019} The magnitude of the transition dipole moment operator is taken to be 4.0 Debye for \Chla\ and 3.4 Debye for \Chlb.\cite{Chan2018} In modelling excitation relaxation in the LHCII monomer it is also necessary to incorporate direct internal conversion of the \Chla* and \Chlb* states. This was done by adding the following non-radiative transition operators to the HEOM/QME
\begin{align}
	\pR_{\LE,\LE}^{\mathrm{NR}} \op{\sigma}_{\LE,\vb{n}}(t) &= -\sum_{n=1}^{N_\LE}\frac{k_{\mathrm{NR},n}}{2}\left\{ \dyad{\LE_n},\op{\sigma}_{\LE,\vb{n}}(t)\right\}\\
	\pR_{\GS,\LE}^{\mathrm{NR}} \op{\sigma}_{\LE,\vb{n}}(t)  &= \sum_{n=1}^{N_\LE} k_{\mathrm{NR},n}\dyad{\GS}{\LE_n}\op{\sigma}_{\LE,\vb{n}}(t)\dyad{\LE_n}{\GS}
\end{align}
and we set $k_{\mathrm{NR},n} = 0.25\ \text{ns}^{-1}$, in line with Cupellini \etal's kinetic model.\cite{Cupellini2020} A similar model has been used previously in quantum master equation based studies of the bacterial LHI/LHII system,\cite{BourneWorster2019} and a justification of this form of the non-radiative transition operator from the Nakajima-Zwanzig equation is given in Appendix \ref{app-icterm}.

In all simulations with the hybrid HEOM/QME method we use an adaptive Short iterative Arnoldi integrator (as described in Ref.~\onlinecite{fay2022simple}) with an error tolerance parameter of $10^{-8}$, and Krylov subspace dimension of $16$. The Nakajima-Zwanzig low temperature and termination corrections described in Ref.~\onlinecite{fay2022simple} were also used in these simulations, and using this scheme the populations of \Chla*, \Chlb* and CT states were found to be converged with an HEOM frequency cut-off parameter of $\Gamma_{\mathrm{c}} = 3 \gamma_{\mathrm{D,Chl^*}}$, using the Matsubara decomposition scheme for the correlation functions. The cut-off parameter for the ET kernels was set to $\Gamma_{\mathrm{c, ET}} = 2\gamma_{\mathrm{D,Chl^*}}$ for the charge separation steps and $\Gamma_{\mathrm{c, ET}} = 0$ for the charge recombination steps. All simulations were run at a temperature of 300 K. These methods are all implemented in the freely available HEOM-lab code,\cite{heom-lab} which was used to perform all simulations presented in this paper.

%with a spectral density given by
%\begin{align}
%	\pazocal{J}_\mathrm{CT}(\omega) &= (1-\alpha) \pazocal{J}_\mathrm{D}(\omega) + \alpha \pazocal{J}_{\mathrm{BO}}(\omega)\\
%	\pazocal{J}_{\mathrm{BO}}(\omega) &= \frac{\lambda}{2}\frac{\gamma\Omega^2\omega}{(\omega^2-\Omega^2)^2+\gamma^2\omega^2}
%\end{align}
%where the first portion is a Debye spectral density with $\omega_{\mathrm{D}} = 30\text{ cm}^{-1}$, which represents the low frequency environment contribution, and $\pazocal{J}_{\mathrm{BO}}(\omega)$ models the high frequency contribution from \ce{C=C} stretches in the lutein molecules. In what follows we examine how the contribution from the 

\subsection{Population dynamics}
\begin{figure}
	\includegraphics[width=0.48\textwidth]{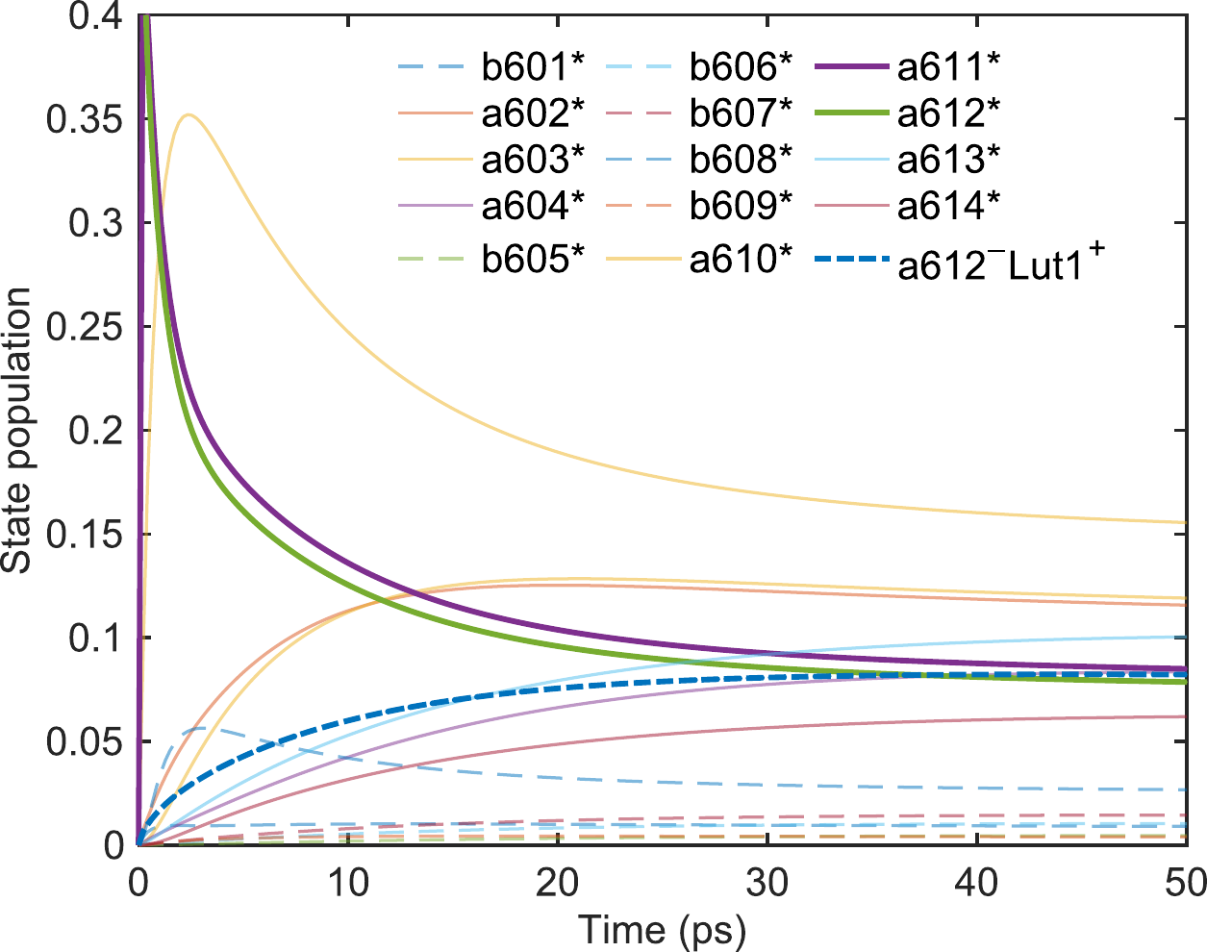}
	\caption{Populations of the LE states and the \chla{612}${}^-$Lut1${}^+$ state calculated with the hybrid HEOM/QME method for an initial excitation localized on \chla{612}, using the model 1 LE Hamiltonian.}\label{site12init-model1-fig}
\end{figure}

As a first application of the hybrid HEOM/QME method to LHCII charge transfer quenching, we have simulated the population dynamics for the \Chla*, \Chlb* and CT states for Model 1, with the initial condition set as an excitation completely localized on \chla{612}. The population dynamics are shown in Fig.~\ref{site12init-model1-fig}. For this initial condition there is an early rapid transfer of population to the strongly coupled \chla{611}* state, as an exciton is formed where the excitation is delocalized between the two \Chla\ sites. This exciton formation is followed by slower excitation energy transfer to the other \Chla* and \Chlb* states, as well as the \chla{612}${}^-$Lut1${}^+$ state. Importantly, formation of the \chla{612}${}^-$Lut1${}^+$ state from the exciton state happens on a comparable time-scale, roughly 10 ps, to energy redistribution between the \Chla* and \Chlb* states  highlighting the need to treat both the CT and exciton dynamics simultaneously. We found that at least $\Gamma_{\mathrm{c}} = 3 \gamma_{\mathrm{D,Chl^*}}$ was needed to converge the population dynamics, with hierarchy termination corrections from Ref.~\onlinecite{fay2022simple}, which corresponds to treating the exciton system-environment coupling up to sixth order in non-Markovian perturbation theory, with partial Markovian eighth order corrections. This illustrates the simple mixed Redfield-F\"orster theories would likely be insufficient for describing the population dynamics of this system, as has been demonstrated previously by Kreisbeck \etal\cite{Kreisbeck2014} This calculation would be intractable with standard HEOM methods due to the very large reorganization energy associated with the charge transfer processes, but with the hybrid HEOM/QME method this calculation is runs in a few minutes on a single CPU.

\begin{figure}[t]
	\includegraphics[width=0.48\textwidth]{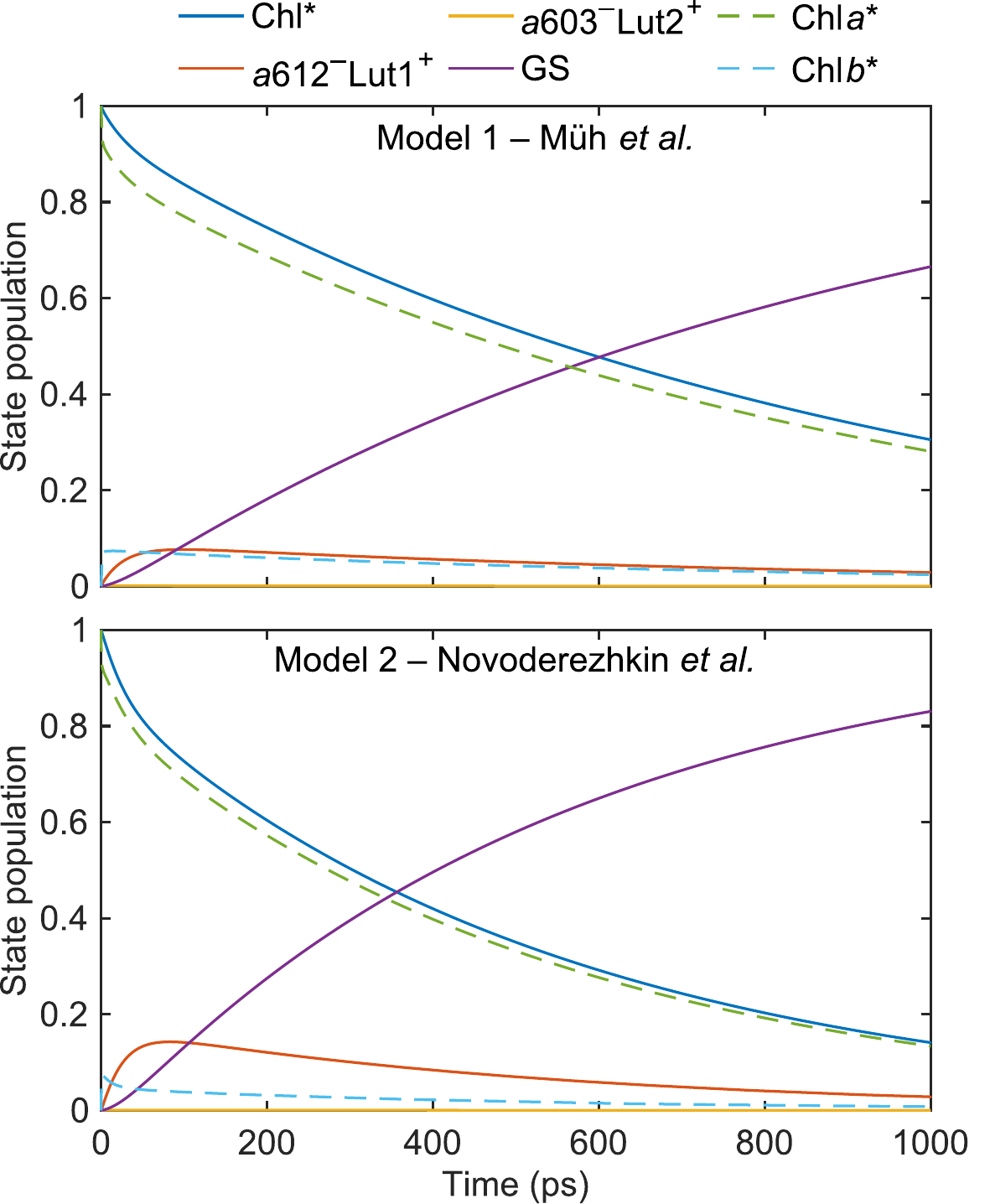}
	\caption{Excited state population dynamics for LHCII with an initial excitation partitioned equally between all eight \Chla* states with no coherences between these states, calcualted with the hybrid HEOM/QME method. Results in the top panel use the model 1 LE Hamiltonian and results in the bottom panel use the model 2 LE Hamiltonian.}\label{chlainit-models-fig}
\end{figure}

In order to compare the kinetic model used by Cupellini \etal\ to our model including the full exciton dynamics, we have also simulated the population dynamics where population is evenly divided between the \Chla* states, with no initial coherences between sites. The total populations of the Chl*, \Chla*, \Chlb*, and the populations of the CT states and GS are shown for both models in Fig.~\ref{chlainit-models-fig}. In both models there is rapid equilibration between the \Chla* and \Chlb* states, occuring on a time scale of a few picoseconds, followed by population transfer to the CT states within ten picoseconds. The \chla{612}${}^-$Lut1${}^+$ state is populated to a much greater extent than the \chla{603}${}^-$Lut2${}^+$ state, because the latter lies nearly 1000 cm${}^{-1}$ above the \Chla* states, whereas the former is nearly degenerate with the \Chla* states. In model 1 we observe a greater extent of population transfer to the \Chlb* states and a lesser population of the \chla{612}${}^-$Lut1${}^+$ state compared to model 2.

%%% I dont think you need to introduce \tau_max. Its follows the same trend and is not as well-defined as \tau_eff
Interestingly, the overall decay of the Chl* population is noticeably faster in model 2 compared to model 1. In order to quantify the excitation lifetime, we fit these population decay curves to a sum of three exponentials (constrained such that $p_{\mathrm{Chl}^*}(t=0) = 1$), and use this to calculate the integrated lifetime $\tau_\mathrm{eff} = \int_0^\infty p_{\mathrm{Chl}^*}(t)  \dd{t}$.
%, and the slowest decay time-scale $\tau_\mathrm{max}$ (the largest of the time constants in the sum of exponentials). The $\tau_\mathrm{eff}$ measure is sensitive to the initial transient population dynamics, whereas $\tau_\mathrm{max}$ is not. 
For model 1 we obtain $\tau_\mathrm{eff} = 0.83$ ns
% and $\tau_\mathrm{max} = 0.88$ ns, 
 and for model 1 we find $\tau_\mathrm{eff} = 0.50$ ns.
 %and $\tau_\mathrm{max} = 0.56$ ns. 
 These lifetime estimates are both closer to the experimental value for the excitation lifetime, 2 ns for LHCII in a membrane, than Cupellini \etal's kinetic model, which predicted a 0.3 ns integrated lifetime. This suggests the importance of treating the exciton dynamics explicitly in modeling charge transfer quenching in light-harvesting complexes. We suspect that a significant source of error in our model, compared to experiment, is the estimate for the diabatic coupling for the charge recombination steps, which affects the lifetime of the CT states, and therefore also strongly influences the CT quenching rate. 

Part of the difference between Cupellini \etal's kinetic model, and our model is that we correctly account for exciton formation in the excited state dynamics, which changes the rate of population transfer from the LE manifold to the CT states. This is because exciton formation modifies the effective coupling matrix element by $\sim 1/\sqrt{2}$ for the charge transfer rate, since charge transfer occurs from the excitonic state $\ket{\psi}\approx ({1}/{\sqrt{2}})(\ket{\chla{611}^*}-\ket{\chla{612}^*})$ (although this picture is complicated by the LE bath reorganization, which reduces the relative coherence between the \chla{611}* and \chla{612}* states to $\sim 0.16$ for model 1 and $\sim 0.20$ for model 2). Furthermore exciton formation lowers the free energy of the initial excitonic state by $\sim 100\text{ cm}^{-1}$ for both models. This means the effective free energy change for the formation of the \chla{612}${}^-$Lut1${}^+$ is approximately $0 \text{ cm}^{-1}$, which is reflected in Fig.~\ref{site12init-model1-fig} where the \chla{611}*, \chla{612}* and \chla{612}${}^-$Lut1${}^+$ states all have approximately equal population by $t = 50$ ps. This means the excitonic state is stabilized relative to the CT state and this decreases the steady state CT population, decreasing the rate of charge separation

The faster quenching in model 2 is primarily due to the larger steady-state population of the \chla{612}${}^-$Lut1${}^+$ which provides the main channel for charge transfer quenching of the Chl* excitations. The larger steady state population of the \chla{612}${}^-$Lut1${}^+$ state can be explained by the differences between the excitonic structure of the two LHCII models. In model 1, the \chla{611}-\chla{612} excitonic state is the third lowest energy state, lying more than $k_{\mathrm{B}} T$ above the lowest lying excitonic state, whereas in model 2 the \chla{611}-\chla{612} excitonic state is the lowest energy state, with the next nearest excitonic state lying about $0.5 k_{\mathrm{B}} T$ above this. This means that in model 2 more of the \Chla* excitations are funnelled into the excitonic state which couples to the quenching state, which leads to a greater extent of CT state formation, and faster Chl* population decay. This illustrates the importance of the excitonic energy funnel in determining quenching efficiency.

\subsection{Role of the excitation energy funnel}
%\begin{figure}
%	\includegraphics[width=0.48\textwidth]{fig10-shiftpops.pdf}
%	\caption{Total Chl* populations for model 1 (left) and model 2 (right) with shifted LE state energies, with excitation initially partitioned equally between the \Chla* states with no initial coherences. The lines correspond to shifts of -500 cm${}^{-1}$ (yellow) through to +500 cm${}^{-1}$ (purple) in steps of 100 cm${}^{-1}$.}\label{shift-pops-fig}
%\end{figure}

\begin{figure}
	\includegraphics[width=0.45\textwidth]{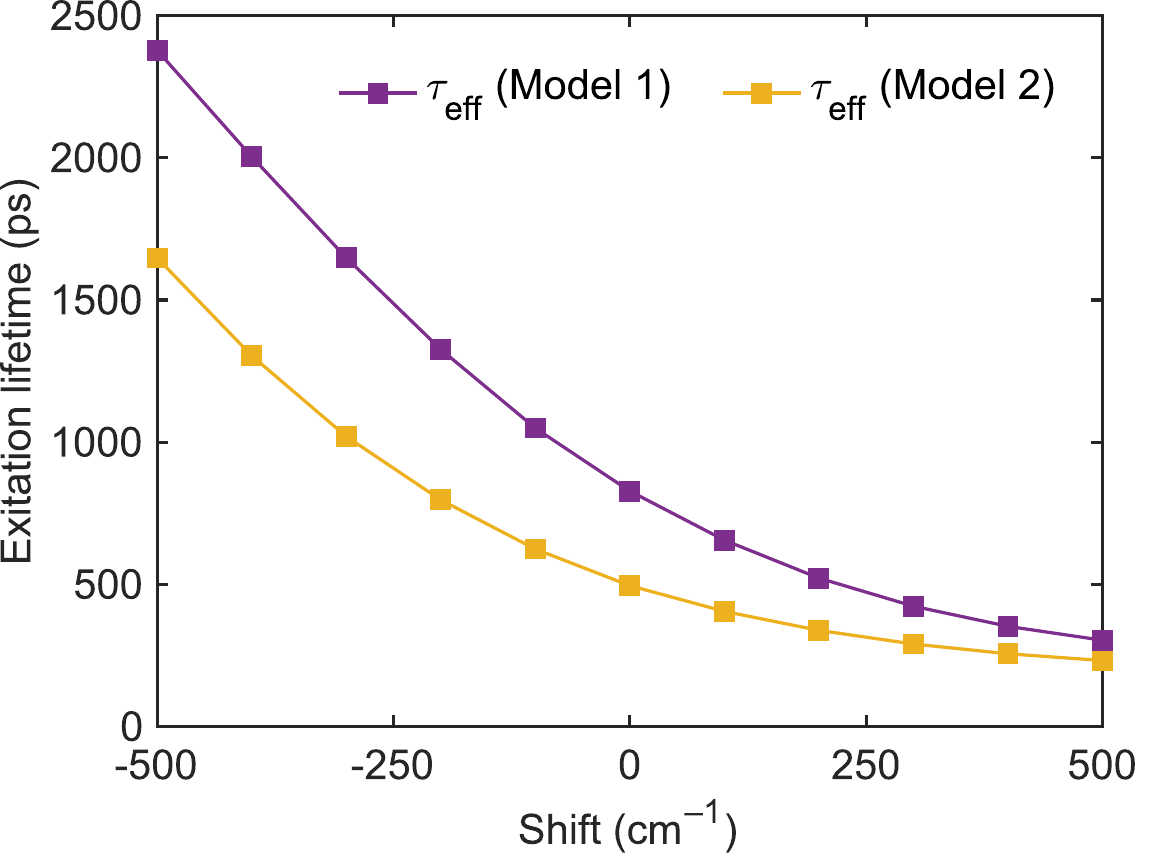}
	\caption{$\tau_\mathrm{eff}$ and $\tau_\mathrm{max}$ for models 1 and 2 as a function of the local excitation energy shift, as described in the text.}\label{shift-taus-fig}
\end{figure}

In order to further explore the effects of the excitonic energy funnel on excitation quenching in LHCII, we have performed simulations on modified versions of the model 1 and model 2 Chl* Hamiltonians. For both models we introduced a shift to the site energies of all states in the LE manifold except the \chla{611} and \chla{612} states, such that the site energies are changed from $E_n$ to $E_n + \delta E$. The energy of the \chla{603}${}^-$Lut2${}^+$ charge transfer is shifted by the same amount. This effectively preserves the \chla{611}-\chla{612} exciton state in the LE manifold, but just shifts its energy relative to the remaining excitonic states. Varying the shift from $-500$ cm${}^{-1}$ to $+500$ cm${}^{-1}$, we have calculated the total Chl* population decay and the integrated lifetime from the population dynamics to quantify the changes to the Chl* excitation lifetime, as shown in Fig.~\ref{shift-taus-fig}.

We see that for modified versions of both model 1 and model 2, shifts in the Chl* site energies can have a very large effect on the population dynamics and excitation lifetime. In particular for model 1 a change in shift from $- 200 \text{ cm}^{-1}$ to $+ 200 \text{ cm}^{-1}$ can change the excitation lifetime by more than a factor of 2. These significant changes in excitation lifetime for modest shifts in the excitation energies suggest a potential mechanism for activation of non-photochemical quenching by changes in the energy funnel in light-harvesting complexes directing excitations towards quenching sites. 
\subsection{Role of the charge transfer spectral density}

We have examined the potential role of nuclear quantum effects in determining the charge transfer quenching lifetime. We modify the spectral density for the charge transfer processes to include an underdamped Brownian oscillator contribution,\cite{Tanimura1989} where the new spectral density is
\begin{align}
	\pazocal{J}_\mathrm{CT}(\omega) &= (1-\alpha) \pazocal{J}_\mathrm{D}(\omega) + \alpha \pazocal{J}_{\mathrm{BO}}(\omega)\\
	\pazocal{J}_{\mathrm{BO}}(\omega) &= \frac{\lambda}{2}\frac{\gamma\Omega^2\omega}{(\omega^2-\Omega^2)^2+\gamma^2\omega^2}
\end{align}
where the first portion is a Debye spectral density with $\omega_{\mathrm{D}} = 30\text{ cm}^{-1}$, which represents the low frequency environment contribution, and $\pazocal{J}_{\mathrm{BO}}(\omega)$ models the high frequency contribution from \ce{C=C} stretches in the lutein molecules.\cite{Balevicius2017} In what follows we have set $\Omega = 1500\text{ cm}^{-1}$ and $\gamma = 50\text{ cm}^{-1}$, and we keep the total reorganization energy for each charge transfer process fixed at the values determined by Cupellini \etal. We have varied the $\alpha$ parameter, which controls the spectral distribution of of the reorganization energy, between 0 and 0.5 for the model 1 exciton Hamiltonian, and for each value of $\alpha$ we have simulated the population dynamics and calculated $\tau_\mathrm{eff}$ and $\tau_\mathrm{max}$, again from an initial condition where all \Chla* states are equally populated with no coherences.

\begin{figure}[t]
	\includegraphics[width=0.48\textwidth]{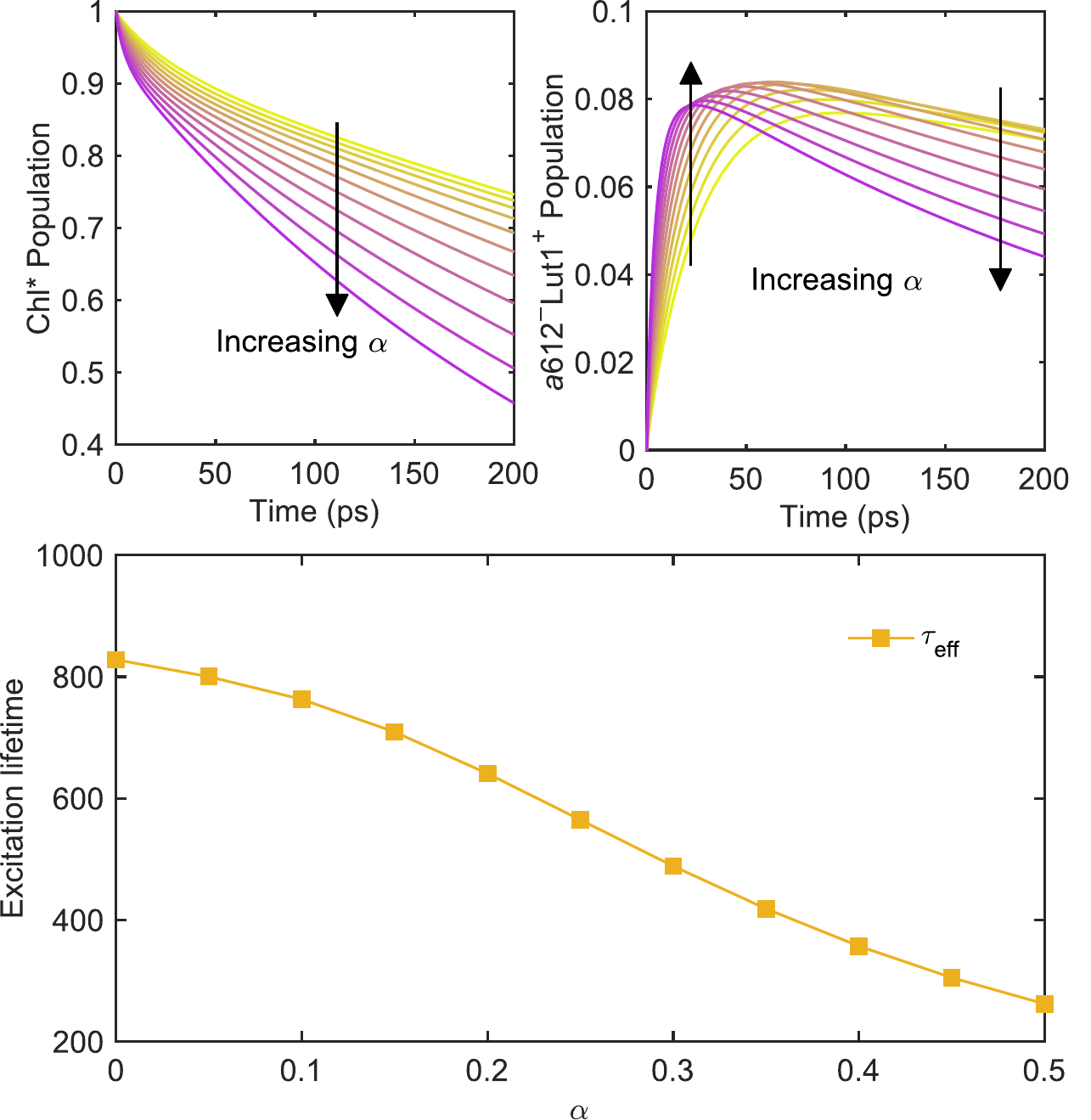}
	\caption{Top left: Chl* population dynamics varying $\alpha$ from 0 (yellow) to 0.5 (purple). Top right: \chla{612}${}^-$Lut1${}^+$ population dynamics varying $\alpha$ (same color scheme as top left). Bottom: excitation lifetime as a function of $\alpha$. All calculations use the model 1 LE Hamiltonian with equally populated \Chla* states with no coherences as the initial condition.}\label{alphaBO-fig}
\end{figure}

The calculated total Chl* and \chla{612}${}^-$Lut1${}^+$  population dynamics are shown in the top panels of Fig.~\ref{alphaBO-fig}. We see that increasing the contribution to the charge transfer spectral density from the underdamped, high frequency Brownian oscillator decreases the lifetime of the Chl* excitations. The increased Brownian oscillator contribution increases the extent of nuclear quantum tunneling, which increases both the rate of charge separation and charge recombination. The effect of increasing the rates of both processes can be seen more clearly in the \chla{612}${}^-$Lut1${}^+$ popuation dynamics. At short times, as $\alpha$ increases, the rate of population transfer from the Chl* manifold to the \chla{612}${}^-$Lut1${}^+$ increases due to an increasing charge separation rate, which transiently increases the quenching rate. At longer times, when the Chl* states and CT states have reached a steady state, increasing $\alpha$ increases the rate of decay of the \chla{612}${}^-$Lut1${}^+$ state, which arises due to the increased charge recombination rate with increasing $\alpha$.

\section{Discussion}\label{sec-disc}

Using the hybrid HEOM/QME method developed in this paper we have been able to explore the charge transfer quenching dynamics in a LHCII monomer. Using previously parametrized models of excitation energy transfer and electron transfer in the LHCII pigment-protein complex, together with some physically motivated assumptions, we can obtain estimates of the excitation lifetime closer to experimental values than those obtained using simple kinetic models.\cite{Cupellini2020} The lifetime estimate however depends strongly on the model of local excitation energies and couplings used to describe the excitation energy transfer preceding charge transfer quenching. This may have interesting implications for the understanding of non-photochemical quenching processes in light harvesting complexes. It is generally assumed that non-photochemical quenching is activated by conformational changes, induced by pH changes or chemical modifications of the protein or bound carotenoids, in protein-pigment complexes moving quenchers into positions where they can efficiently coupled to chromophores,\cite{Park2019,Short2022,Chrysafoudi2021,Daskalakis2020,Ruban2021,Lapillo2020,Goss2015} thereby increasing the rate of charge transfer quenching and/or excitation energy transfer quenching. The work here highlights the potential role of the excitation energy funnel in CT quenching, and we suggest that non-photochemical quenching could also be activated by changes in the site energies and couplings within the chromophore excited state manifold that funnel excitations towards quenching sites. These energy and coupling shifts could also be induced by conformational changes in the protein-pigment complex, and could occur simultaneously with the conformational changes that move quenchers closer to chromophores. In this way, the energy funnelling mechanism could work cooperatively with quenching activation by movement of quenchers. Interestingly it has recently been found in one computational study that pH changes, which are believed to play a role in activating non-photochemical quenching, can create modest shifts in the Chl excitation energies in LHCII.\cite{Maity2021} 

Our exploration of the charge transfer spectral density has also highlighted the importance of understanding all details of charge transfer processes in light harvesting complexes, including the details of recombination processes and nuclear quantum effects. One particularly important factor for predicting charge transfer quenching rates is the rate of charge recombination from the CT state back to the electronic ground state of the system, which is strongly dependent on nuclear quantum effects as well as the energetics of charge recombination and the diabatic coupling strength. This is because the reverse electron transfer is often deep in the Marcus inverted regime, for example in LHCII the Lut-\Chla\ recombination processes have a free energy change of $-\Delta G \approx 3 \lambda$, where nuclear tunnelling effects play a decisive role in determining electron transfer rates.\cite{Lawrence2019a} Although in our study of LHCII we have assumed that the electron transfers can be treated with a spin-boson mapping,\cite{Blumberger2015} the theory developed here does not rely on this assumption. The correlation functions needed to evaluate the ET kernels, $G_{AB}^C(t)$, can be evaluated using various approximations including anharmonicity in the potential energy surfaces.\cite{Sparpaglione1988,Rabani1999,Egorov1999,Tong2020a,Lawrence2018,Lawrence2020a} For example $G_{AB}^C(t)$ could be evaluated classically,\cite{Sparpaglione1988} or using analytic continuation together with path integral methods to incorporate anharmonic nuclear quantum effects.\cite{Lawrence2018}

\section{Concluding remarks}\label{sec-conc}

In this paper we have outlined how to rigorously combine the hierarchical equations of motion method with quantum master equations in both the strong system-bath coupling and weak system-bath coupling limits, to model simultaneous excitation energy transfer and charge transfer in protein-pigment complexes. The hybrid HEOM/QME approach is based the application of Zwanzig projection, to derive a system of equation for a hierarchy of auxiliary density operators for the various electronic state manifolds. This method has been tested against numerically exact results for an excitonic dimer coupled to a charge transfer state, where it was found to yield accurate population and coherence dynamics across a range of excitation energy transfer regimes. We then applied the method to study charge transfer quenching in LHCII - a process suspected to play an important role in photoprotection in plants.

Using the hybrid HEOM/QME approach we have been able to study the interplay of excitation energy transfer and charge transfer quenching in a realistic model of LHCII. Our results highlight the importance of the excitation energy funnel in determining quenching efficiency in protein-pigment complexes, as well as the role of the CT state recombination rate when back electron transfer to reform LE states from CT states occurs at an appreciable rate, as is the case in LHCII. We expect that the energy funnel mechanism could play a role in activation of non-photochemical quenching in many systems, and that it could occur co-operatively with other NPQ activation mechanisms.

The parameters needed in the hybrid HEOM/QME model can all be obtained using well-established methods, for example exciton Hamiltonians can be fit based on spectroscopic data,\cite{Muh2010,Novoderezhkin2011} or using QM/MM simulations of protein-pigment complexes.\cite{Cignoni2022} Likewise electron transfer model parameters can be obtained using molecular dynamics simulations.\cite{Blumberger2015,Tong2020a} By combining existing computational tools with the hybrid HEOM/QME method, currently implemented in the freely available Matlab code HEOM-lab,\cite{heom-lab} it may be possible to shed light on the precise mechanisms that produce non-photochemical quenching in photosynthetic organisms,\cite{Goss2015} for example how chemical modifications of carotenoids in the xanthophyll cycle activates NPQ in LHCII and related proteins like LHCX1.\cite{Park2019,Short2022} For this reason we anticipate the method will become a useful tool in studying non-photochemical quenching and reaction center processes in photosynthetic systems. 

\section*{Acknowledgements}
T. P. F and D.T.L. were supported by the U.S. Department of Energy, Office of Science, Basic Energy Sciences, CPIMS Program Early Career Research Program under Award DE-FOA0002019.

\appendix

\section{The radiative decay term}\label{app-rad}

We evaluate the radiative coupling term in the hybrid HEOM/QME in much the same way as the ET coupling term. Here we assume that only the $ \LE $ states and the $\GS$ are connected by the dipole moment operator, which means we only have to evaluate $\pR_{\GS,\LE}^{\mathrm{R}}$ and $\pR_{\LE,\LE}^{\mathrm{R}}$. Starting with $\pR_{\GS,\LE}^{\mathrm{R}}$ we can insert the dipole coupling term into the second order kernel and directly evaluate the radiative coupling term as
\begin{align}
	\begin{split}
		\pR_{\GS,\LE}^{\mathrm{R}} &= {\frac{ 1}{2\hbar\pazocal{V}_0\varepsilon_0}}\sum_{\vb*{k},p} \int_0^\infty \!\!\!\dd{t}\omega_{\vb*{k}} \\
		&\times\bigg( 
		\pL_{{\vb*{k}p}}^{\mathrm{L}} e^{\pL_{0,\sys} t} \pL_{{\vb*{k}p}}^{\mathrm{R}} \ev{a_{\vb*{k}p}(0)a_{\vb*{k}p}^\dag(t)}_{{\bath_{\EM}}}\\
		&+\pL_{{\vb*{k}p}}^{\mathrm{R}} e^{\pL_{0,\sys} t} \pL_{{\vb*{k}p}}^{\mathrm{L}} \ev{a_{\vb*{k}p}(0)a_{\vb*{k}p}^\dag(t)}_{{\bath_{\EM}}}^*\bigg)
	\end{split}
\end{align}
where we have assumed that we can approximate $\pL_0 + \pazocal{V} \approx \pL_{0,\sys} $ in evaluating this term. The operators $\pL_{{\vb*{k}p}}^{\mathrm{L/R}}$ are given by
\begin{align}
	\pL_{{\vb*{k}p}}^{\mathrm{L}} \op{\sigma} &= i \op{\Pi}_{\GS}(\op{\vb*{\mu}}\cdot \vb*{e}_{\vb*{k}p} ) \op{\Pi}_{\LE}\op{\sigma} \\
	\pL_{{\vb*{k}p}}^{\mathrm{R}} \op{\sigma} &= -i \op{\sigma}\op{\Pi}_{\LE}(\op{\vb*{\mu}}\cdot \vb*{e}_{\vb*{k}p} )^\dag\op{\Pi}_{\GS},
\end{align}
and the EM field operator correlation function can be evaluated as
\begin{align}
	\ev{a_{\vb*{k}p}(0)a_{\vb*{k}p}^\dag(t)}_{{\bath_{\EM}}} = \Tr_{{\bath_{\EM}}}[\op{a}_{\vb*{k}p}(0)\op{a}_{\vb*{k}p}^\dag(t)\op{\rho}_{{\bath_{\EM}}}]= e^{i \omega_{\vb*{k}} t}.
\end{align}
In the limit of a large cavity volume for the EM field $\pazocal{V}_0$, we can replace the sum over cavity modes $\vb*{k}$ with an integral as
\begin{align}
	\frac{1}{\pazocal{V}_0}\sum_{\vb*{k}} \to \frac{1}{(2\pi)^3} \int\dd{\vb*{k}}.
\end{align}
After evaluating the angular part of the integral we arrive at
\begin{align}
	\begin{split}
		&\pR_{\GS,\LE}^{\mathrm{R}} = {\frac{ 1}{6\hbar\varepsilon_0\pi^2}}\sum_{\alpha=x,y,z}\int_0^\infty \!\!\!\dd{k} \int_0^\infty \!\!\!\dd{t}c_0 k^3\\
		&\times\bigg( \pL_{\alpha}^{\mathrm{L}} e^{\pL_{0,\sys}^{\LE,\GS} t} \pL_{\alpha}^{\mathrm{R}} e^{i c_0 k  t }+\pL_{\alpha}^{\mathrm{R}} e^{\pL_{0,\sys}^{\GS,\LE} t} \pL_{\alpha}^{\mathrm{L}}  e^{-i c_0 k  t }\bigg),
	\end{split}
\end{align}
then we insert the spectral resolution of $\pL_{0,\sys} $ acting on the $\GS$-$\LE$ coherences ($\dyad{\GS}{\LE_n}$), and note that all the eigenvalues of this operator are purely imaginary, so we can write these as $\pL_{0,\sys}^{\GS,\LE}  = \pazocal{S}_{\GS,\LE}^{0,\sys} (i\Omega_{\GS,\LE}) ({\pazocal{S}_{\GS,\LE}^{0,\sys}})^{-1}$ and $\pL_{0,\sys}^{\LE,\GS}  = \pazocal{S}_{\LE,\GS}^{0,\sys} (-i\Omega_{\LE,\GS}) ({\pazocal{S}_{\LE,\GS}^{0,\sys}})^{-1}$, where $\Omega_{\GS,\LE}$ and $\Omega_{\LE,\GS}$ are diagonal matrices with real positive-valued entries. We then evaluate the time integral, noting the imaginary part vanishes, and change variables in the $k$ integral to $\omega = c_0 k$ to give
\begin{align}
	\begin{split}
		\pR_{\GS,\LE}^{\mathrm{R}} &= {\frac{ 1}{6 \hbar \varepsilon_0 c_0^3 \pi}}\sum_{\alpha=x,y,z}\int_0^\infty \!\!\!\dd{\omega} \omega^3\\
		&\times\bigg( \pL_{\alpha}^{\mathrm{L}} \pazocal{S}_{\LE,\GS}^{0,\sys} \delta(\Omega_{\LE,\GS}-\omega) ({\pazocal{S}_{\LE,\GS}^{0,\sys}})^{-1} \pL_{\alpha}^{\mathrm{R}} \\
		&+\pL_{\alpha}^{\mathrm{R}} \pazocal{S}_{\GS,\LE}^{0,\sys} \delta(\Omega_{\GS,\LE}-\omega) ({\pazocal{S}_{\GS,\LE}^{0,\sys}})^{-1} \pL_{\alpha}^{\mathrm{L}}  \bigg).
	\end{split}
\end{align}
Integrating over $\omega$ then yields in the expressions given in Eq.~\eqref{R-rad-GSLE-eq}. These steps can be repeated for $\pR_{\LE,\LE}^{\mathrm{R}}$ to obtain Eq.\eqref{R-rad-LELE-eq}, where additionally we discard imaginary terms (which correspond to Lamb shifts).

\section{Simplification of the dimer model baths}\label{app-dimerbaths}

In this appendix we describe how the three-bath model for the dimer-CT model can be reduced to a two bath model to speed up exact HEOM calculations. Suppose we have a system coupled to a set of baths, $j=1,...,N$, with identical frequency distributions, but different reorganization energies, i.e. $\pazocal{J}_j(\omega) = \eta_j \pazocal{J}_0(\omega)$ with $\eta_j = \lambda_j / \lambda_0$, as in the dimer CT model considered here. We can write the coupling term for bath modes with frequency $\omega_{j\alpha} = \omega_{0\alpha}$ as
\begin{align}
	\op{H}_{\sys\bath,\alpha} = c_\alpha\op{\vb*{q}}_{\alpha}\cdot  \vb*{\eta}^{1/2}\op{\vb*{V}}
\end{align}
where $c_\alpha = c_{0\alpha}$, $[\op{\vb*{q}}_\alpha]_j = \op{q}_{j\alpha}$, $[\op{\vb*{V}}]_j = \op{V}_j$, and $\vb*{\eta}$ is a diagonal matrix of the values of $\eta_j$. We can insert an orthogonal matrix $\vb{S}$, to re-write this as
\begin{align}
	\op{H}_{\sys\bath,\alpha} = c_\alpha\op{\vb*{q}}_{\alpha}^\mathsf{T}\vb{S}^\mathsf{T} \vb{S}  \vb*{\eta}^{1/2}\op{\vb*{V}} = c_\alpha \op{\tilde{\vb*{q}}}\cdot\op{\tilde{\vb*{V}}}.
\end{align}
We can then re-write the Hamiltonian in terms of a new set of uncorrelated baths with mode displacements $\op{\tilde{{q}}}_{j\alpha}$, and coupling operators $\op{\tilde{V}}_j$. If we choose $\vb{S}$ such that one of the new system bath coupling operators is just proportional to the identity operator $\op{\tilde{V}}_1 \propto \op{1}$, then we eliminate coupling between one of the baths and the system, and therefore reduce the complexity of the problem. For $\kappa = 1$ we can find such a transformation as follows. First we write $\vb*{\eta}^{1/2}\op{\vb*{V}}$ as
\begin{align}
	\vb*{\eta}^{1/2}\op{\vb*{V}} =\left(\begin{matrix}
		\dyad{\LE_1} \\
		\dyad{\LE_2} + \kappa\dyad{\CT} \\
		\sqrt{\eta} \dyad{\CT}
	\end{matrix}\right)
\end{align} 
where $\lambda_0 = \lambda_{\LE}$ and $\eta = \sqrt{\lambda_{\CT}/\lambda_{\LE}}$. We then set $\vb{S}$ as
\begin{align}
	\vb{S}=\left(\begin{matrix}
		\frac{1}{\sqrt{2}} & \frac{1}{\sqrt{2}} & 0 \\
		\frac{1}{\sqrt{2}} & -\frac{1}{\sqrt{2}} & 0 \\
		0 & 0 & 1
	\end{matrix}\right)
\end{align} 
which gives $\op{\tilde{\vb*{V}}}$ as
\begin{align}
\op{\tilde{\vb*{V}}} =\left(\begin{matrix}
	\frac{1}{\sqrt{2}}\op{1} \\
	\frac{1}{\sqrt{2}}(\dyad{\LE_1}-\dyad{\LE_2}  - \dyad{\CT}) \\
	\sqrt{\eta} \dyad{\CT}
\end{matrix}\right).
\end{align}
We see that the system coupling operator for the new bath 1 is proportional to an identity operator, so coupling to this bath does not affect the system dynamics and it can eliminated. 

For the $\kappa = 0$ case we set
\begin{align}
	\vb{S}=\left(\begin{matrix}
		\frac{1}{\sqrt{2 + \eta^{-1}}} & \frac{1}{\sqrt{2 + \eta^-1}} &  \frac{1}{\sqrt{\eta(2+\eta^{-1})}}\\
		\frac{1}{\sqrt{2}} & -\frac{1}{\sqrt{2}} & 0 \\
	-\frac{1}{\sqrt{2+4\eta}} & -\frac{1}{\sqrt{2+4\eta}} & \frac{2\sqrt{\eta}}{\sqrt{2+4\eta}}
	\end{matrix}\right)
\end{align} 
which gives 
$\op{\tilde{\vb*{V}}}$ as
\begin{align}
	\op{\tilde{\vb*{V}}} =\left(\begin{matrix}
		\frac{1}{\sqrt{2+\eta^{-1}}}\op{1} \\
		\frac{1}{\sqrt{2}}(\dyad{\LE_1}-\dyad{\LE_2}) \\
		\frac{1}{\sqrt{2+4\eta}}\left(2\eta \dyad{\CT}-\dyad{\LE_1} -\dyad{\LE_2}\right) 
	\end{matrix}\right).
\end{align}
Again bath 1 can be eliminated because its coupling operator is proportional to an identity operator. We note that only the top row of $\vb{S}$ is uniquely defined for any $\kappa$, by requiring that $\op{\tilde{V}}_1 \propto \op{1}$, and the choice for the other rows, and other coupling operators, is not unique.

Finally we note that we can scale the reorganization energy of a given bath by $\alpha$ if we also scale the corresponding coupling operator by $1/\sqrt{\alpha}$, because the coupling coefficients are proportional to $\sqrt{\lambda_j}$, $c_{j\alpha}\propto \sqrt{\lambda_j}$, without changing the Hamiltonian. The choice does not affect the exact dynamics, but it does affect how the the hierarchy is truncated with our reorganization energy weighted cut-off scheme. For the calculations we set $\lambda_2 = \lambda_{\LE}$ and $\lambda_3 = \lambda_{\CT}$, which means the coupling operator $\op{\tilde{V}}_3$ given above scaled down by $1/\sqrt{\eta}$.

\section{The internal conversion term}\label{app-icterm}

In this appendix we present a brief justification of the Lindblad form for the internal conversion term used to model direct non-radiative transitions from the \ce{Chl^*} states to the ground state in LHCII. The argument is essentially the same as that used to derive the radiative and ET coupling Hamiltonians. We start by assuming that the internal conversion coupling is described by a Hamiltonian of the form
\begin{align}
	\op{H}_{\LE,\GS} = \sum_{n=1}^{N_\LE} \left(\op{X}_n \dyad{\GS}{\LE_n}+ \op{X}_n^\dag \dyad{\LE_n}{\GS}\right)
\end{align}
where we assume the $\op{X}_n$ operator acts on degrees of freedom other than the $\bath_{\LE}$ degrees of freedom, and we also assume that these operators commute at all times. We also assume that the thermal average of $\op{X}_n$ is zero. The non-adiabatic coupling between the LE states and the ground state depends primarily on local vibrational modes of the chromophore, so it is reasonable to assume that these operators commute. 

We can now evaluate the second order Markovian Nakajima-Zwanzig relaxation operator for internal conversion with the projection operator given in main text. We will further approximate the reference Liouvillian as $\pL_0 + \pazocal{V} \approx -\frac{i}{\hbar}[\op{\Pi}_{\LE}\bar{E}_{\LE},\ \cdot\ ]\otimes \pazocal{I}_{\mathrm{ado}}$, where $\bar{E}_{\LE}$ is the average LE state energy, an approximation that is justified because the mean energy difference between the LE states and the ground-state is much larger than energy differences within the LE manifold. From this it is straightforward to obtain the relaxation superoperator as
\begin{align}
	\begin{split}
	\pR_{\LE,\LE}^{\mathrm{NR}} \op{\sigma}&= \sum_{n=1}^{N_\LE} \int_0^\infty \!\!\!\dd{t} \bigg(\ev{X_n^\dag(t)X_n(0)}_\bath e^{i \bar{E}_{\LE} t/\hbar} \dyad{\LE_n}\op{\sigma} \\
	&+\ev{X_n^\dag(t)X_n(0)}_\bath^* e^{-i \bar{E}_{\LE} t/\hbar} \op{\sigma}\dyad{\LE_n}\bigg).
	\end{split}
\end{align}
Ignoring Lamb shift terms that originate from the imaginary parts of the $\ev{X_n^\dag(t)X_n(0)}_\bath$ correlation functions, this reduces to
\begin{align}
	\pR_{\LE,\LE}^{\mathrm{NR}} \op{\sigma} &= -\sum_{n=1}^{N_\LE}\frac{k_{\mathrm{NR},n}}{2}\left\{ \dyad{\LE_n},\op{\sigma}\right\}
\end{align}
where $k_{\mathrm{NR},n}$ is the non-radiative decay rate for internal conversion of state $\LE_n$. The corresponding term in the equation of motion for $\op{\sigma}_{\GS,\vb{n}}(t)$ is
\begin{align}
	\pR_{\GS,\LE}^{\mathrm{NR}} \op{\sigma} &= \sum_{n=1}^{N_\LE}{k_{\mathrm{NR},n}} \dyad{\GS}{\LE_n}\op{\sigma}\dyad{\LE_n}{\GS}. 
\end{align}
The reverse internal conversion rate is related to the forward rate by $k_{\mathrm{NR},n}^{\mathrm{back}} = e^{-\beta \bar{E}_\LE}k_{\mathrm{NR},n} $, but because $\beta\bar{E}_{\LE}$ is typically very large, we can safely ignore the back reaction terms.

\section{Additional details of the LHCII models}\label{lhcii-model-app}

In this appendix we list the model parameters used in our simulations of the LHCII monomer. Firstly the LE system Hamiltonians for the two LHCII models are
\begin{widetext}
\begin{align}
\vb{H}_{\LE,\sys} = (E_\mathrm{LE} + \lambda_{\mathrm{Chl}^*})\vb{1}+ \left( 	\begin{blockarray}{*{14}{c}}
		635 & 36 & -5 & -3 & 1 & -2 & -3 & 3 & 4 & -5 & 20 & 2 & -8 & 2 \\
		36 & 70 & 15 & 6 & 0 & 5 & 6 & -6 & -24 & -5 & 1 & 8 & -2 & 0 \\
		-5 & 15 & 80 & -1 & 0 & -4 & 6 & 4 & 72 & 7 & -1 & 1 & 1 & -5 \\
		-3 & 6 & -1 & 140 & 4 & 71 & 24 & -4 & -2 & 0 & -3 & 3 & 2 & -3 \\
		1 & 0 & 0 & 4 & 775 & 9 & -4 & -4 & 0 & 1 & 1 & -2 & -1 & 0 \\
		-2 & 5 & -4 & 71 & 9 & 615 & 16 & -5 & 2 & 0 & -2 & 2 & 2 & -2 \\
		-3 & 6 & 6 & 24 & -4 & 16 & 525 & -4 & -5 & 1 & -2 & 3 & 3 & -3 \\
		3 & -6 & 4 & -4 & -4 & -5 & -4 & 395 & 24 & 43 & 5 & -1 & -2 & 1 \\
		4 & -24 & 72 & -2 & 0 & 2 & -5 & 24 & 855 & -2 & 4 & -1 & -2 & 2 \\
		-5 & -5 & 7 & 0 & 1 & 0 & 1 & 43 & -2 & 0 & -26 & 13 & 6 & -1 \\
		20 & 1 & -1 & -3 & 1 & -2 & -2 & 5 & 4 & -26 & 150 & 99 & -3 & 1 \\
		2 & 8 & 1 & 3 & -2 & 2 & 3 & -1 & -1 & 13 & 99 & 180 & 0 & 0 \\
		-8 & -2 & 1 & 2 & -1 & 2 & 3 & -2 & -2 & 6 & -3 & 0 & 90 & -36 \\
		2 & 0 & -5 & -3 & 0 & -2 & -3 & 1 & 2 & -1 & 1 & 0 & -36 & 200 
	\end{blockarray} 
\right)\text{cm}^{-1}
\end{align}
for model 1, and
\begin{align}
	&\vb{H}_{\LE,\sys} = (E_\mathrm{LE} + \lambda_{\mathrm{D,Chl}^*})\vb{1}\nonumber\\
	&+ \left( 	\begin{blockarray}{*{14}{c}}
		816.00 & 49.64 & -5.89 & -2.51 & 0.77 & -1.87 & -2.49 & 2.78 & 3.79 & -5.95 & 24.89 & 9.13 & -10.79 & 3.59 \\
		49.64 & 84.00 & 38.11 & 6.42 & -0.71 & 5.60 & 7.13 & -5.84 & -19.25 & -11.39 & 9.69 & 15.83 & -4.96 & 0.69 \\
		-5.89 & 38.11 & 214.00 & -3.28 & 1.13 & -8.89 & 1.23 & 6.72 & 96.66 & 12.97 & -2.70 & -0.76 & 2.68 & -6.70 \\
		-2.51 & 6.42 & -3.28 & 387.00 & 3.35 & 104.56 & 35.93 & -2.76 & -7.28 & -4.18 & -3.80 & 4.67 & 2.12 & -3.42 \\
		0.77 & -0.71 & 1.13 & 3.35 & 606.00 & 29.71 & -4.47 & -5.13 & -0.77 & 1.61 & 1.33 & -2.85 & -1.40 & 0.37 \\
		-1.87 & 5.60 & -8.89 & 104.56 & 29.71 & 777.00 & 59.38 & -4.99 & -0.16 & -3.28 & -2.52 & 3.10 & 1.47 & -2.16 \\
		-2.49 & 7.13 & 1.23 & 35.93 & -4.47 & 59.38 & 641.00 & -4.43 & -11.99 & -0.14 & -2.78 & 3.07 & 2.20 & -3.25 \\
		2.78 & -5.84 & 6.72 & -2.76 & -5.13 & -4.99 & -4.43 & 688.00 & 36.07 & 61.97 & 4.35 & -1.08 & -2.01 & 1.30 \\
		3.79 & -19.25 & 96.66 & -7.28 & -0.77 & -0.16 & -11.99 & 36.07 & 648.00 & 3.86 & 4.30 & -2.57 & -2.92 & 2.33 \\
		-5.95 & -11.39 & 12.97 & -4.18 & 1.61 & -3.28 & -0.14 & 61.97 & 3.86 & 0.00 & -24.96 & 23.10 & 7.21 & -1.55 \\
		24.89 & 9.69 & -2.70 & -3.80 & 1.33 & -2.52 & -2.78 & 4.35 & 4.30 & -24.96 & 39.00 & 126.92 & -6.15 & 4.55 \\
		9.13 & 15.83 & -0.76 & 4.67 & -2.85 & 3.10 & 3.07 & -1.08 & -2.57 & 23.10 & 126.92 & 21.00 & -0.47 & -0.18 \\
		-10.79 & -4.96 & 2.68 & 2.12 & -1.40 & 1.47 & 2.20 & -2.01 & -2.92 & 7.21 & -6.15 & -0.47 & 101.00 & -50.22 \\
		3.59 & 0.69 & -6.70 & -3.42 & 0.37 & -2.16 & -3.25 & 1.30 & 2.33 & -1.55 & 4.55 & -0.18 & -50.22 & 187.00 
	\end{blockarray} 
	\right)\text{cm}^{-1}
\end{align}
\end{widetext}
for model 2. The columns/rows correspond to the states in the following order: \textit{b}601*, \textit{a}602*, \textit{a}603*, \textit{a}604*, \textit{b}605*, \textit{b}606*, \textit{b}607*, \textit{b}608*, \textit{b}609*, \textit{a}610*, \textit{a}611*, \textit{a}612*, \textit{a}613*, \textit{a}614*. In these Hamiltonians we incorporate the reorganization energy contribution of the LE baths into the system Hamiltonian matrices, and as such diagonal elements correspond to vertical excitation energies in the absence of LE coupling, and the diagonal element minus $\lambda_{\mathrm{D},\mathrm{Chl}^*}$ is the free energy of that LE state in the absence of inter LE  state coupling. The CT state system Hamiltonians are given by
\begin{align}
	\op{H}_{\CT_1,\sys} = (E_{\chla{612}^*}+\Delta G_{\CT1})\dyad{\CT_1} \\
	\op{H}_{\CT_2,\sys} = (E_{\chla{603}^*}+\Delta G_{\CT2})\dyad{\CT_2}
\end{align}
where and $E_{\chla{612}^*}$ and $E_{\chla{603}^*}$ are the diagonal elements of $\vb{H}_{\LE,\sys}$ corresponding the states \chla{612}* and \chla{603}* respectively. Here $\CT_1$ is the \chla{612}${}^-$Lut1${}^+$ state and $\CT_2$ is the \chla{603}${}^-$Lut2${}^+$ state. Finally $\op{H}_{\GS,\sys} = 0$ by definition. The remaining model parameters are listed in Tab.~\ref{lhcii-params-tab}.

The $G_{AB}^B(t)$ and $G_{BA}^B(t)$ functions were evaluated using\cite{Fay2018,Fay2021c}
\begin{align}
	G_{AB}^B(t) &= G_{BA}^B(t)^* = \exp(\zeta_{AB}(t)+i\Delta\epsilon_{AB} t /\hbar) \\
	\begin{split}
	\zeta_{AB}(t) &= -\int_0^\infty \dd{\omega} \frac{\pazocal{J}_{AB}(\omega)}{\omega^2} \\
	&\times \left[\coth(\frac{\beta\hbar\omega}{2})(1-\cos(\omega t))+i\sin(\omega t)\right]
	\end{split}
\end{align}
 where $\pazocal{J}_{AB}(\omega)$ is the $\bath_{\CT}$ spectral density associated with the $A\to B$ charge transfer, and $\Delta \epsilon_{AB}$ is the free energy change excluding $\bath_{\LE}$ contributions (i.e. the free energy change with $\op{B}_{n,r} = 0$). In order to evaluate the integrals over these functions, each component of the spectral density (the Debye and Brownian Oscillator components) was discretized into 512 frequencies using a Gauss-Legendre quadrature for the function $(1/(4\pi\lambda))\pazocal{J}_{\mathrm{D}/\mathrm{BO}}(\omega)/\omega$,\cite{Lawrence2020a} and using this the $\zeta_{AB}(t)$ function was evaluated and the required numerical integrals were evaluated using the trapezium rule. The integrals were evaluated up to $t_{\mathrm{max}} = 23.5$ fs, discretized into 1000 time points for the CT-LE correlation functions and 14104 time points for the much more oscillatory CT-GS correlation functions. The same expressions and methodology were used to evaluate the ET kernel for the dimer model in Sec.~\ref{sec-dimer}.
\begin{table}[b]
	\begin{tabular}{cc}
		\hline
		Parameter & Value (cm${}^{-1}$ unless specified otherwise) \\
		\hline
		$\Delta G_{\CT1}$ & $-$82 \\
		$\lambda_{\CT1}$ & 5405 \\
		$V_{\CT1}$ & 240 \\
		$\Delta G_{\CT2}$ & 951\\
		$\lambda_{\CT2}$ & 5052 \\
		$V_{\CT2}$ & 279 \\
		$\omega_{\mathrm{D},\CT}$ & 30 \\
		$\Omega$ & 1500 \\
		$\gamma$ & 50 \\
		$E_{\LE}$ (Model 1) & 14780 \\
		$E_{\LE}$ (Model 2) & 15073 \\
		$\lambda_{\mathrm{D},\mathrm{Chl}^*}$ & 220\\
		$\omega_{\mathrm{D},\mathrm{Chl}^*}$ & 353.6777 \\
		$\mu_{\mathrm{\Chla}}$ & 4.0 D \\
		$\mu_{\mathrm{\Chlb}}$ & 3.4 D \\
		$k_{\mathrm{NR}}$ & 0.25 ns${}^{-1}$\\
		\hline
	\end{tabular}
\caption{Parameters used for the LHCII monomer CT quenching models.}\label{lhcii-params-tab}
\end{table}

\section*{References}
\bibliography{bibliography.bib}

\end{document}